\documentclass[prb,twocolumn,superscriptaddress,showpacs]{revtex4}
\usepackage[dvips]{graphicx}
\usepackage{epsfig}
\usepackage{latexsym}
\usepackage{bm}

\renewcommand{\ni}{{\noindent}}
\newcommand{\be}{\begin{equation}}
\newcommand{\ee}{\end{equation}}
\newcommand{\bea}{\begin{eqnarray}} 
\newcommand{\eea}{\end{eqnarray}}
\newcommand{\nn}{\nonumber} 
\newcommand{\bk}{{\bf k}}
\newcommand{\la}{\langle}
\newcommand{\ra}{\rangle} 

\newcommand{\br}{{\bf{r}}} 
\newcommand{\brp}{{\bf{r}^\prime}} 
\newcommand{\bq}{{\bf{q}}}
\newcommand{\hx}{\hat{\bf x}} 
\newcommand{\ha}{\hat{\alpha}} 
\newcommand{\hb}{\hat{\beta}} 

\newcommand{\bS}{{\bf S}} 
\newcommand{\bR}{{\bf R}}

\newcommand{\up}{\uparrow} 
\newcommand{\dn}{\downarrow}
\newcommand{\tDelta}{{\tilde\Delta}}
\newcommand{\tmu}{{\tilde\mu}}

\newcommand{\sig}{\sigma}
\newcommand{\sigb}{{\bar{\sigma}}}
\newcommand{\sigp}{{\sigma^\prime}}
\newcommand{\sigpb}{{\bar{\sigma}^\prime}}

\def\AA{\stackrel{_{~\circ}}{\textstyle A}}

\begin{document}

\title{High Tc Superconductors: A Variational Theory of the
Superconducting State}
\author{Arun Paramekanti}
\affiliation{Tata Institute of Fundamental Research, Mumbai 400 005, India}
\affiliation{Department of Physics and Kavli Institute for
Theoretical Physics, University of California, Santa Barbara, 
California 93106--4030}
\author{Mohit Randeria}
\affiliation{Tata Institute of Fundamental Research, Mumbai 400 005, India}
\affiliation{Department of Physics, University of Illinois at 
Urbana-Champaign, Urbana, IL 61801}
\author{Nandini Trivedi}
\affiliation{Tata Institute of Fundamental Research, Mumbai 400 005, India}
\affiliation{Department of Physics, University of Illinois at 
Urbana-Champaign, Urbana, IL 61801}
\begin{abstract}
\vspace{0.1cm}
We use a variational approach to gain insight into the strongly correlated 
d-wave superconducting state of the high Tc cuprates at T=0. 
We show that strong correlations lead to qualitatively different trends in 
pairing and phase coherence: the pairing scale decreases monotonically 
with hole doping while the SC order parameter shows a non-monotonic 
dome. We obtain detailed results for the doping-dependence of a large 
number of experimentally observable quantities, including the
chemical potential, coherence length,
momentum distribution, nodal quasiparticle weight 
and dispersion, incoherent features in photoemission spectra, 
optical spectral weight and superfluid density. Most of our results are in 
remarkable quantitative agreement with existing data and some
of our predictions, first reported in Phys. Rev. Lett. {\bf 87}, 217002 (2001),
have been recently verified.
\typeout{polish abstract}
\end{abstract}
\date{\today}
\maketitle


\section{Introduction}

In this paper our main goal is to understand the superconducting 
ground state and low energy excitations of the high Tc cuprates, in 
particular, their doping dependence as they evolve from a 
Fermi liquid state on the overdoped side
towards a Mott insulator at half-filling. 
Toward this end, we examine in detail the properties of 
superconducting wavefunctions in which double occupancy is strongly 
suppressed by short-range Coulomb interactions. 

In the past seventeen years since
the discovery of high Tc superconductivity (SC) 
in the cuprates \cite{hightc86},
a lot of theoretical effort has gone into trying to understand 
SC as an instability from a non-superconducting state.
There are three possible routes to such an attack and each
has its own strengths and limitations. First, one may approach the SC state from
the overdoped side, where the normal state is a well-understood Fermi liquid.
However, the diagrammatic methods used in such an approach are not adequate 
for addressing the most interesting underdoped region in
the vicinity of the Mott insulator. Second, one might hope to
examine the SC instability from the near-optimal normal state,
except that this normal state is highly abnormal and the breakdown of 
Fermi-liquid behavior remains one of the biggest open questions in the field. 
The third approach is to enter the SC state as a doping-driven instability from
the Mott insulator. While this approach has seen considerable theoretical 
progress, much of the discussion is complicated by various broken 
symmetries and competing instabilities in lightly doped
Mott insulators.

Here we take a rather different approach, in which we do not view 
superconductivity as an instability from any non-SC state, but rather study the
SC state in and of itself. After all, the main reason for
interest in the cuprates comes from their {\em superconductivity},
and not from other possible orders, which may well exist in specific
materials in limited doping regimes.
Thus it is very important to theoretically
understand the SC state in all its details, particularly capturing both
the BCS-like behavior on the overdoped side and the non-BCS aspects,
like the large spectral gap but
low Tc and superfluid density, 
on the underdoped side. The resulting insights could also 
help in characterizing the anomalous normal states which are 
obtained upon destroying the SC order.

We choose to work within a two-dimensional, single-band approach with
strong local electron-electron interactions, described by the large $U$
Hubbard model, first advocated by Anderson \cite{anderson87} for the 
cuprate superconductors. Our goal is to see how much of the
physics of the cuprates can be captured within this framework.
To the extent that this approach proves inadequate, one may need to go beyond
it and include either long-range Coulomb interactions,
additional bands, inter-layer effects, or even
phonons. The success of our approach reported here suggest to us that,
at least for the SC state properties studied,
one does not need to explicitly include these additional degrees of freedom.

The key technical challenge is to 
treat the effect of strong correlations in a controlled manner.
We have chosen to deal with this using Gutzwiller
wavefunctions and using the variational Monte Carlo method
to evaluate various expectation values building on pioneering
work by several authors \cite{anderson87,fczhang,gros88,shiba88}. 

We now summarize our main results;
this also serves as an outline of the remainder of the paper.
Some of these results were first reported in a Letter
\cite{paramekanti01}.

\ni$\bullet$ 
We introduce in Sec.~IV our wavefunction which is a fully
projected d-wave BCS state, in which all configurations with
doubly occupied sites are first eliminated, and the
effects of the finite Coulomb $U$ are then
built in via a canonical transformation
described in Sec.~III.

\ni$\bullet$ 
Using a variational calculation, we obtain in Sec.~V.B
the following $T=0$ phase diagram: a Fermi liquid metal for 
hole concentrations $x > x_c \approx 0.35$, 
a strongly correlated d-wave SC for $0 < x < x_c$, 
and a spin-liquid Mott insulator at $x=0$. 

\ni$\bullet$ 
The pairing, as characterized by our variational
parameter $\Delta_{\rm var}(x)$, is a monotonically decreasing
function of hole doping $x$, largest at $x=0$ and vanishing
beyond $x_c$; see Fig.~1(a). In marked contrast, the SC order parameter
$\Phi(x)$ shows a nonmonotonic doping dependence; see Fig.~3(a).

\ni$\bullet$ 
The nonmonotonic SC order parameter naturally gives rise to
the notion of optimal doping $x \simeq 0.2$ at which 
superconductivity is the strongest.
We explain in detail in Sec.~V.B this nonmonotonic
behavior and, in particular, how strong correlations -- and not
a competing order -- lead to a suppression of
superconductivity as $x \to 0$ despite the presence of strong pairing.

\ni$\bullet$ 
We predict a nonmonotonic doping dependence for the SC coherence length, 
$\xi_{\rm sc}$, which diverges both as $x\to 0^+$ and as
$x\to x_c^-$, but is small, of order few lattice spacings 
at optimal doping; see Fig.~3(b) and Sec.~V.C.

\ni$\bullet$
We study the momentum distribution $n(\bk)$ and its doping
dependence in Sec.~VI, and find that the ``Fermi surface'' derived from 
$n(\bk)$ is consistent with angle-resolved photoemission spectroscopy
(ARPES) experiments and very similar to the non-interacting
band-theory result.

\ni$\bullet$
Using the singularities of the moments of the electronic 
spectral function, we characterize in detail the low-lying
excitations of the SC state in Secs.~VII and VIII.
We obtain, for the first time, the doping dependence of the
coherent weight $Z$ and Fermi velocity $v_{_F}$ of nodal quasiparticles (QP)
and make predictions for the nodal QP self-energy. 
Remarkably, $Z$ vanishes as $x \to 0$, however $v_{_F}$ is essentially
doping independent and finite as $x \to 0$.
Our predictions \cite{paramekanti01} for the magnitudes and doping
dependence of the nodal $Z$ and $v_{_F}$ have been verified by
new ARPES experiments as discussed in Sec.~VIII.

\ni$\bullet$
We demonstrate, using moments of the electronic spectral function, that 
strong correlations lead to large incoherent spectral weight which
is distributed over a large energy scale.
Specifically, we relate our variational parameter $\Delta_{\rm var}(x)$
to an incoherent energy scale at $\bk = (\pi,0)$ in Sec.~IX.
This motivates us to compare $\Delta_{\rm var}$ with the incoherent 
$(\pi,0)$ ``hump'' scale in ARPES. As seen from Fig.~8, once we
scale $\Delta_{\rm var}$ to agree with the data at one doping value, 
we find excellent agreement between the two for all doping levels.

\ni$\bullet$ 
We compute in Sec.~X the total optical spectral
weight $D_{\rm tot}$ and the low frequency optical spectral
weight or Drude weight $D_{\rm low}$; see Fig.~10(a). 
The Drude weight, which vanishes with underdoping, 
is in good quantitative agreement with optics experiments.  

\ni$\bullet$ 
We predict that the Drude weight
$D_{\rm low} \sim Z$ nodal QP weight, in the entire SC regime, 
which could be tested by comparing optics and ARPES experiments. 

\ni$\bullet$ 
We use the calculation of $D_{\rm low}$ to obtain an upper bound
on the superfluid density, leading to the conclusion that
the superfluid density vanishes as $x\to 0$, consistent with experiments.
The underdoped regime thus has strong pairing but very small phase stiffness
leading to a pairing pseudogap above $T_c$.

This is first time that such a wealth of information
has been obtained on correlated SC wavefunctions which permits useful
comparison with and predictions for a variety of experiments. 
Further, as we will emphasize later in the text,
many qualitative features of our results in the underdoped region, 
such as the incoherence 
in the spectral function and the doping dependence of quantities like
the SC order parameter, nodal QP weight, and Drude weight
are mainly the consequence of the projection, which imposes the
no double-occupancy constraint, rather than of other aspects of the
wavefunction or details of the Hamiltonian. 

Three appendices contain technical details. Appendix A describes
the canonical transformation and its effect on various operators 
used throughout the text. Appendix B contains details about
the Monte Carlo method used and various checks on the program.
Finally, Appendix C is a self contained summary of the slave boson
mean field theory calculations with which we compare our variational
results throughout the text.

\section{Comparison with other approaches:}

In a field with a literature as large as the high Tc superconductors
it is important to try and make a clear comparison of our approach and its 
results with those of other approaches. In this section we will
briefly endeavor to do this.

First, a few remarks about the choice of Hamiltonian:
as indicated in the Introduction, we wish to explore 
strongly correlated one band systems, since this is clearly
a minimal description of the cuprates.
We have chosen to work with the strong
coupling Hubbard model and find that the results are more reasonable than
those for the tJ model as evidenced, e.g., in the comparison of the momentum
distributions of the two models (see Fig.~6(a)) 
and well-known differences in sum rules \cite{eskes94}.
These differences arise in part because in the tJ model
certain $t^2/U$ terms (superexchange) are retained while others
(three-site hops) discarded. More importantly, the Coulomb $U$ is
treated asymmetrically in the tJ model: it is large
but finite in the order $J \sim t^2/U$ term retained in
the Hamiltonian but set to infinity in so far as the upper cutoff
and other operators are concerned as discussed further in Sec.~III.
However, these differences may well be matters of detail.

The more important point is that no variational calculation can ever
{\em prove} that the Hubbard or tJ model has a SC ground state in some
given doping and parameter range. While some numerical studies
hint at a SC ground state in the tJ model \cite{sorella}, such
studies, which attempt to improve upon a variational wavefunction, are 
naturally biased by the choice of their starting state. More direct 
numerical attacks \cite{assaad,zhang,white} on these models have been 
unable to provide unambiguous 
answers to this question for both technical (fermion sign problem and small
system sizes) and physical (competition between various ordered states at 
low doping) reasons.

The exact ground state very likely depends sensitively on details
of the Hamiltonian, for instance, presence of small ring-exchange terms.
We are thus less interested here in the exact ground state of a particular
microscopic Hamiltonian, and more in the properties of strongly
correlated superconducting wavefunctions.
Motivated by our work \cite{paramekanti01}, Laughlin \cite{laughlin02} has
recently inverted the problem to find the Hamiltonian for which a
certain correlated SC wavefunction is the exact ground state.

In any case, it is clear that one needs to study Hamiltonians
like the Hubbard model in which the largest energy scale is the
on-site Coulomb correlation. The key question then is how one treats this,
or indeed any, strongly interacting 2D Hamiltonian.
The two approaches explored in detail in the literature are
variational wave functions and slave bosons.
Variational
Gutzwiller wavefunctions were introduced in the first paper
of Anderson \cite{anderson87} and extensively studied by the Zurich and
Tokyo groups \cite{fczhang,gros88,shiba88,heeb94,ogata96,ogata00} 
and others using both
exact Monte Carlo methods and Gutzwiller approximations.
The primary focus of these studies was the ground state energetics
of various competing phases, and in fact d-wave superconductivity was
predicted by an early variational calculation \cite{gros88}.

Our work builds upon these earlier studies. The new aspects of our
work are the following:
\noindent
(1) We propose a wavefunction in which we first fully project out
doubly-occupied sites and then back off from $U = \infty$ using
a canonical transformation.
\noindent
(2) We focus {\em not} on the energy, which can certainly be further improved by
additional short-range Jastrow factors, but rather on various experimentally
observable quantities. 
\noindent
(3) We exploit sum rules to write frequency moments
of dynamical correlation functions as equal-time correlators which
can be calculated within our method. 
\noindent
(4) We exploit the singularities
of moments to extract information about the important low-lying excitations:
the nodal quasiparticles. 
\noindent
(5) Through the study of moments, we also extract 
information about incoherent features in electronic spectral functions,
which are an integral part of strongly correlated systems and have not
been studied much theoretically. 

Through out the text (and in Appendix C) we
compare our results with those obtained within
slave boson method mean field theory (SBMFT) 
\cite{bza87,kotliar88}.
The chief advantage of this approach is its simplicity
but there are important questions about its validity,
especially the approximations of treating 
the no-double occupancy constraint in an average manner.
The resulting answers for the overall phase diagram in the
doping-temperature plane are very suggestive
\cite{nagaosa92,fukuyama92}.
However there are major
problems: the fluctuations about the mean field, which are necessary
to include in order to impose the constraint reliably, are described by
a strongly coupled gauge theory over which one 
has no control, in general \cite{dhlee}.
In view of this, the very language of spinons and holons, which are
the natural excitations at the mean field level, is suspect
since these are actually strongly coupled degrees of freedom rather than
forming a ``quasiparticle'' basis in which to solve the problem.
The same conclusion is also reached by comparing SBMFT results
with the corresponding results from our variational calculation
where the constraint is imposed exactly; see Sec.~VIII and Appendix C.
In our approach we always work in terms of the physical electron 
coordinates.

\section{Model}

A minimal model for strongly correlated electrons on a
lattice is the single-band Hubbard model defined by the Hamiltonian
\be
{\cal H}={\cal K} + {\cal H}_{\rm int} \nn \\
=\sum_{\bk,\sig} \epsilon(\bk)c_{\bk\sig}^{\dag} c_{\bk\sig}
+ U \sum_{\br} n_{\br\uparrow} n_{\br\downarrow}
\label{hubbard}
\ee
The kinetic energy ${\cal K}$ is governed by
the free electron dispersion $\epsilon(\bk)$,
and ${\cal H}_{\rm int}$ describes the local Coulomb repulsion
between electrons. $n_{\br\sig} = c_{\br\sig}^{\dag} c_{\br\sig}$
is the electron number operator.

We write ${\cal K} = - \sum_{\br,\brp,\sig} t_{\br\brp}
c_{\br\sig}^{\dag}c_{\brp\sig}$ in real space, and set the hopping
$t_{\br\brp} = t$ for nearest neighbors, $t_{\br\brp} = -t^\prime$
for next-nearest neighbors and $t_{\br\brp} = 0$ for other
$(\br,\brp)$ on a two-dimensional ($2D$) square lattice. This
leads to the dispersion $\epsilon(\bk) = -2t\left(\cos k_x + \cos
k_y \right) + 4t^\prime \cos k_x \cos k_y$. The need to include a
$t^\prime > 0$ term in the dispersion is suggested by modeling of
ARPES data \cite{norman95,kim98} and electronic structure calculations
\cite{andersen01}.

We will focus on the strong correlation regime of this model,
defined by $U \gg t,t^\prime$ and low hole doping $x$,
where the number density of electrons $\langle n \rangle = 1 - x$.
Thus $x = 0$
corresponds to half-filling, with one electron per site. To make
quantitative comparison with the cuprates, we choose
representative values: $t=300 meV$, $t^\prime=t/4$, and $U=12 t$.
The values of $t$  and $t^\prime$ are obtained from band theory
estimates; $t$ sets the scale of the bandwidth, while the choice
of (the sign and value of) $t^\prime$ controls mainly the shape or
topology of the ``Fermi surface'' as shown in below in Sec.~VI. 
A non-zero $t^\prime$ also ensures that we break bipartite symmetry,
which can be important for certain properties \cite{vison}.
The Coulomb $U = 12t$ is chosen such that the nearest neighbor
antiferromagnetic exchange coupling $J= 4 t^2/U = 100 meV$,
consistent with values obtained from inelastic light scattering
\cite{lyons88} and neutron scattering experiments
\cite{keimer89,aeppli01} on the cuprates.
There are no more adjustable parameters once $t, t^\prime$ and $U$
are fixed.

The Hilbert space of the electrons described by the Hubbard model
has four states at each site: $\vert 0 \ra$, $\vert\up\ra$,
$\vert\dn\ra$, and $\vert\up\dn\ra$. Many-body configurations can then be
labeled by the total number of doubly occupied sites in the
lattice ${\cal D} = \sum_{\br} n_{\br\up} n_{\br\dn}$. In the
large $U$ limit we focus on the low-energy subspace with no
doubly occupied sites: ${\cal D} = 0$. Towards this end we use the
canonical transformation, originally due to Kohn \cite{kohn64}, and
subsequently used in the derivation of the 
tJ model \cite{gros87} from the large $U$ Hubbard model. 
We discuss this transformation in some detail since it is used to
define our variational wavefunction as
discussed in the next section, and it will also be important in understanding
the differences between Hubbard and tJ results.

The unitary transformation \cite{kohn64,gros87,yoshioka88}
$\exp(iS)$ is defined so that the transformed Hamiltonian
$\tilde{{\cal H}} \equiv \exp(iS){\cal H}\exp(-iS)$ has no matrix
elements connecting sectors with different double occupancy ${\cal
D}$. For large $U$, we can determine $S$ perturbatively in
$(t/U)$, such that the off-diagonal matrix elements of
$\tilde{\cal H}$ between different ${\cal D}$-sectors are
eliminated order by order in $(t/U)$.

Following Ref.~\cite{yoshioka88}, we
write the kinetic energy as ${\cal K}=
{\cal K}_0+{\cal K}_{-1}+{\cal K}_{+1}$, where ${\cal K}_n$ acting
on a state increases ${\cal D}$ by $n$. Thus, ${\cal K}_0$ conserves
${\cal D}$, ${\cal K}_{-1}$ leads to ${\cal D}\to{\cal D}-1$ and
${\cal K}_{+1}$ leads to ${\cal D}\to{\cal D}+1$. Defining the hole
number operator $h_{\br\sig} = (1-n_{\br\sig})$ and
$\sigb = - \sig$, we find
\bea
{\cal K}_0 &=& - \sum_{\br,\brp,\sig} t_{\br\brp} \big[ n_{\br\sigb}
c_{\br\sig}^{\dag}c_{\brp\sig} n_{\brp\sigb}
+ h_{\br\sigb} c_{\br\sig}^{\dag}c_{\brp\sig} h_{\brp\sigb}
\big]\nn\\
{\cal K}_{+1} &=& - \sum_{\br,\brp,\sig} t_{\br\brp} n_{\br\sigb}
c_{\br\sig}^{\dag}c_{\brp\sig} h_{\brp\sigb} \nn \\
{\cal K}_{-1} &=& - \sum_{\br,\brp,\sig} t_{\br\brp} h_{\br\sigb}
c_{\br\sig}^{\dag}c_{\brp\sig} n_{\brp\sigb}
\label{Kndefn}
\eea
The resulting transformation
to ${\cal O}(t/U)^2$ is \cite{yoshioka88}
\bea
i S &\equiv& i S^{\left[1\right]} + i S^{\left[2\right]} \nn \\
&=& \frac{1}{U} \left( {\cal K}_{+1} - {\cal K}_{-1} \right)
+ \frac{1}{U^2} \left( \left[{\cal K}_{+1},{\cal K}_{0}\right]
+ \left[{\cal K}_{-1},{\cal K}_{0}\right] \right)
\label{exprS}
\eea

Using the expression for $S$ to ${\cal O}(t/U)$ the 
transformed Hamiltonian in the sector with ${\cal D}=0$ is
given by
\bea
\tilde{{\cal H}} = {\cal K}_0 -
\sum_{\br,\brp,\bR,\sig\sigp} {{t_{\br\bR} t_{\bR\brp}}\over U} 
&\big(& h_{\br\sigb} c_{\br\sig}^{\dag} c_{\bR\sig} n_{\bR\sigb} \nn\\
&\times& c_{\bR\sigp}^{\dag}
c_{\brp\sigp} h_{\brp\sigpb}\big). 
\label{tildeH}
\eea
Here we have retained all terms to order $t^2/U$. These are
of two kinds: 
(1) Exchange or interaction terms of the form
$\bS_\br\cdot\bS_\brp$ or $n_{\br} n_{\brp}$, 
where $S^\alpha_{\br} = 
{1 \over 2}c_{\br\sig}^{\dag}\tau^\alpha_{\sig,\sigp} c_{\br\sigp}$
with $\tau^\alpha$ the Pauli matrices ($\alpha = x,y,z$). These terms
arise when $\br = \brp$ in Eq.~(\ref{tildeH}). 
(2) 3-site hopping terms of the form 
$h_{\br\sigb}c_{\br\sig}^\dag c_{\bR\sig} n_{\bR\sigb}
c_{\bR\sigp}^{\dag} c_{\brp\sigp} h_{\brp\sigpb}$, which arise
when $\br \neq \brp$.

The  tJ model may be obtained from
the above model as follows: (i) Keep $U/t \gg 1$ but finite in
$\tilde{\cal H}$ leading to 2-site interaction terms of ${\cal
O}(J)$ but drop the 3-site terms which are also ${\cal O}(J)$, 
where $J \equiv 4 t^2/U$, and
(ii) Take $U/t \to \infty$ in the canonical transformation
$\exp(iS)$ for all operators {\em other than} the Hamiltonian, so
that these are {\em not} transformed. Clearly, the above simplifications
are not consistent for the Hubbard model at any $U/t
\gg 1$. However, we may view the tJ model, derived in this manner,
as an interesting model in its own right, capturing some of the 
nontrivial strong correlation physics of the large-$U$ Hubbard model. With the
constraint on the Hilbert space, $\sum_{\br\sig} n_{\br\sig} \leq
1$ at each $\br$, the tJ model is defined by the Hamiltonian 
\bea
{\cal H}_{\rm tJ}=&-&\sum_{\br,\brp,\sig} t_{\br\brp}
c_{\br\sig}^{\dag} c_{\brp\sig} \nn\\
&+& \frac{1}{2} \sum_{\br\brp}
J_{\br\brp} \left( \bS_\br\cdot\bS_\brp - \frac{1}{4} n_{\br}
n_{\brp}\right),
\label{HtJ} 
\eea 
where $J_{\br\brp} = 4 t_{\br\brp}^2/U$.

We will compare below our results for the large $U$
Hubbard model with the corresponding results for the tJ model in order
to understand the importance of the canonical transformation on
various operators and of the inclusion of
the 3-site terms in $\tilde{\cal H}$. In addition we will also
compare our variational results with slave-boson
mean-field theory (SBMFT) for the tJ model.

\section{The variational wavefunction}

Our variational ansatz for the ground state of the high Tc superconductors
is the Gutzwiller projected BCS wavefunction
\be
\vert \Psi_0 \rangle = \exp(-iS) {\cal P} \vert \Psi_{\rm BCS} \rangle.
\label{wavefn}
\ee
We now describe each of the three terms in the above equation.
$\vert \Psi_{\rm BCS} \rangle = \left(\sum_\bk \varphi(\bk)
c_{\bk\uparrow}^{\dag}c_{-\bk\downarrow}^{\dag}\right)^{N/2} \vert 0
\rangle$
is the $N$-electron d-wave BCS wave function \cite{BCS_footnote} with
$\varphi(\bk) = v_\bk/u_\bk
= \Delta_\bk/[\xi_\bk + \sqrt{\xi_\bk^2 + \Delta_\bk^2}]$.
The two variational parameters $\mu_{\rm var}$ and $\Delta_{\rm var}$
enter the pair wavefunction $\varphi(\bk)$ through
$\xi_\bk = \epsilon(\bk) - \mu_{\rm var}$ and
$\Delta_\bk = \Delta_{\rm var}\left(\cos k_x - \cos k_y\right)/2$.
%
%
Since the wavefunction is dimensionless, it is important to realize
that the actual variational parameters are the dimensionless $u_k$ and $v_k$,
or equivalently the dimensionless quantities: 
$\tmu_{\rm var} = \mu_{\rm var}/t$ and $\tDelta_{\rm var} = \Delta_{\rm var}/t$.

The numerical calculations (whose details are described in Appendix B)
are done in real space. The wavefunction is written as a Slater
determinant of pairs
\be
\langle \{\br_i\},\{\brp_j\} \vert \Psi_{\rm BCS} \rangle
= {\rm Det} \vert\vert \varphi(\br_i - \brp_j) \vert\vert,
\label{det}
\ee 
where 
$\{\br_i\}$  and $\{\brp_j\}$ are the coordinates of the 
spin-up and down electrons respectively, and
$\varphi(\br_i - \brp_j)$ is the Fourier transform of $\varphi(\bk)$.

We focus on the d-wave state in part
motivated by the experimental evidence in the cuprates, but also
because very early variational calculations \cite{gros88,shiba88} predicted 
that the d-wave SC state is energetically the most favorable over 
a large range of hole doping. It is also straightforward to see, at
a mean field level, that large $U$ Hubbard and tJ models should favor
d-wave superconductivity \cite{dwave_mft} with
superexchange $J$ mediating the pairing.

The effect of strong correlations comes in through the
Gutzwiller projection operator 
${\cal P} \equiv \Pi_{\br}(1-n_{\br\up}n_{\br\dn})$ which eliminates
all doubly occupied sites from $\vert\Psi_{\rm BCS} \rangle$ as
would be appropriate for $U/t = \infty$. We back off from infinite $U$
using the unitary operator $\exp(-iS)$ defined above, which
builds in the effects of double occupancy perturbatively
in powers of $t/U$ without introducing any new variational
parameters. For the most part we will need $S$ to ${\cal
O}(t/U)$, so that $S^{\left[1\right]}$ will suffice. However, in some
calculations, we will need to keep the $(t/U)^2$ corrections 
arising from $S^{\left[2\right]}$. 

To understand the role of $\exp(-iS)$, note that
for any operator ${\cal Q}$,
\bea
\la\Psi_0\vert {\cal Q}\vert \Psi_0\ra
&=& \la\Psi_{\rm BCS}\vert {\cal P} \tilde{\cal Q} {\cal P} \vert\Psi_{\rm BCS}
\ra,
\label{transformed_op}
\eea
where $\tilde{\cal Q} \equiv \exp(iS) {\cal Q} \exp(-iS)$.
The fully projected wavefunction
${\cal P}\vert\Psi_{\rm BCS}\ra$ is an appropriate ansatz for the
ground state of the canonically transformed Hamiltonian $\tilde{\cal H}$
in the sector with ${\cal D}=0$. Thus, incorporating
the $\exp(-iS)$ factor in the wavefunction is entirely equivalent
to canonically transforming {\it all} operators ${\cal Q} \to
\tilde{\cal Q} = \exp(iS) {\cal Q} \exp(-iS)$. This has important
consequences, some of which were noted previously in Ref.~\cite{eskes94},
and which will be discussed in detail below.

We emphasize that our wavefunction Eq.~(\ref{wavefn}) is {\it not} the same as
the {\it partially projected} Gutzwiller wavefunction
$\prod_{\br} \left[1-(1-g) n_{\br\uparrow} n_{\br\downarrow} \right]
\vert \Psi_{\rm BCS} \rangle$ with an additional variational
parameter $0 < g < 1$.
Such partially projected states have recently been reexamined 
by Laughlin and dubbed ``gossamer superconductors'' \cite{laughlin02}.
The advantage of such an approach is that by exploiting
the invertability of partial projectors one can identify a Hamiltonian
for which such a state is the exact ground state.
The differences between partial projection and
our approach are most apparent at half filling ($x=0$).
As we will show, our wavefunction Eq.~(\ref{wavefn}) describes 
a Mott insulator with a vanishing low energy optical (Drude) weight at $x=0$.
In contrast, the partially projected Gutzwiller wavefunction has 
non-zero Drude weight at $x=0$ and continues
to be superconducting at half-filling \cite{laughlin02}. 

The inability of {\em partially projected}
states to describe Mott insulators 
at half-filling and sum-rule problems 
for such states are well known \cite{millis91}. 
As shown in Ref.~\cite{millis91} a calculation of the optical conductivity 
based on partially projected states leads to the (unphysical) result
$\int_{0^{+}}^\infty \,d\omega{\rm Re}\sigma(\omega)=0$ even 
though ${\rm Re}\sigma(\omega) \ne 0$ for $\omega > 0$ for
Hubbard-like Hamiltonian. The important property of the Hamiltonian
used for this result is that the vector potential couples only to
the kinetic energy which is quadratic in the electron operators.
It seems likely that the ``gossamer'' Hamiltonian is not of this
type and may avoid the sum rule problem.

In this work we have preferred to use $\exp(-iS){\cal P}$, 
rather than a partial
projection, to build in the effects of a large but finite $U$. This 
permits us to obtain a Mott insulator at $x=0$ and avoid the
unphysical optical conductivity problem for the Hubbard Hamiltonian.

\subsection{Optimal variational parameters}

The first step in any variational calculation is to 
minimize the ground state energy $\la{\cal H}\ra
\equiv \la\Psi_0\vert {\cal H} \vert\Psi_0\ra/\la\Psi_0\vert\Psi_0\ra$
at each doping value $x$.
This determines the optimal values of the (dimensionless) variational parameters 
$\tDelta_{\rm var}$ and $\tmu_{\rm var}$ as functions of the hole-doping $x$, 
From now on $\la {\cal Q} \ra$ will denote
the expectation value of an operator ${\cal Q}$
in the normalized, optimal state $\vert\Psi_0\ra$.
For a two-dimensional $N$-particle system the required expectation values can be written
as $2N$-dimensional multiple integrals which are calculated using
standard Monte Carlo techniques \cite{ceperley}, the 
technical details of which are given in Appendix B.

\begin{figure}
\begin{center}
\vskip-2mm
\hspace*{1mm}
\epsfig{file=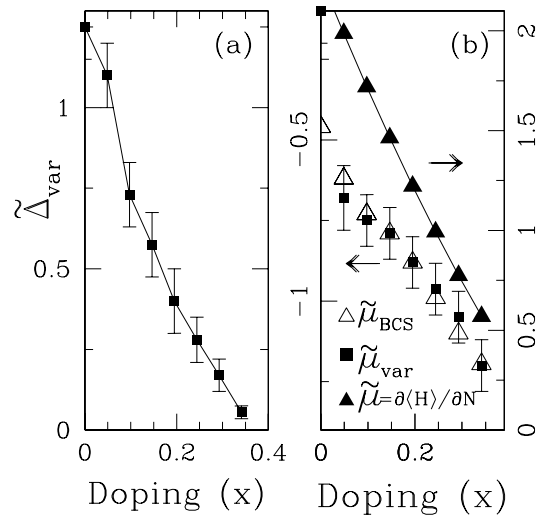,height=3.0in,width=3.0in,angle=0}
\vskip1mm
\caption{(a) Doping dependence of the dimensionless
variational parameter $\tDelta_{\rm var}$ (filled squares). 
%
%
(b) Doping dependence of the dimensionless variational parameter 
$\tmu_{\rm var}(x)$ (filled squares) and the ``BCS value'' $\tmu_{\rm BCS}(x)$ (open triangles) 
defined in the text, both plotted on the scale given on the left-hand y-axis, 
The physical chemical potential (in units of t)
$\tmu = \mu/t$ where $\mu = \partial\la{\cal H}\ra/\partial N$ (filled triangles) 
is plotted on the right-hand y-axis scale.  
}
\label{figdeltamu}
\end{center}
\end{figure}

The optimal $\tDelta_{\rm var}(x)$ is plotted as a function of doping
in Fig.~\ref{figdeltamu}(a). We find that it is finite at $x=0$, and
decreases monotonically with increasing $x$, vanishing beyond a
critical $x=x_c \approx 0.35$. 
%
%
We will show in the next Section that, in marked contrast to
simple BCS theory, $\tDelta_{\rm var}(x)$ is {\em not} the SC order parameter.
Its relationship to the spectral gap will be clarified in Section IX; for now it is simply a 
variational parameter that characterizes pairing in the
internal wave function.

For $x > x_c \approx 0.35$, $\tDelta_{\rm var}(x)=0$, there is no pairing
and the system has a Fermi liquid ground state, which is expected at 
sufficiently large doping \cite{engelbrecht}. 
At $x=x_c$ there is a transition to
a d-wave SC, with the superexchange interaction leading to pairing.
We have found numerically that the value of $x_c$ 
is weakly dependent on $J$ and $t$ for a range of values around the chosen
ones. A similar result is also obtained from slave-boson mean-field theory 
in Appendix C. A crude estimate for $x_c$ may be obtained as follows.
With increasing hole doping, a given electron has fewer neighboring electrons
to pair with, leading to an effective interaction 
$J_{\rm eff} = J (1- 4 x)$, where
the factor of $4$ is the coordination number on the $2D$ square lattice.
The vanishing of $J_{\rm eff}$ determines $x_c =0.25$, which is
both independent of $J$ and in reasonable
agreement with variational estimate $x_c \approx 0.35$, given the crudeness
of the argument.


The optimal value of the second variational parameter $\tmu_{\rm var}(x)$
is plotted in Fig.~\ref{figdeltamu}(b). 
It is important to distinguish this quantity
from the chemical potential of the system
$\mu = \partial\la{\cal H}\ra /\partial N$. As seen
from Fig.~\ref{figdeltamu}(b), $\tmu(x) = \mu/t$ and $\tmu_{\rm var}(x)$ have
quite different magnitudes and doping dependences, in marked contrast
with simple BCS theory, where the two would have been identical. 

To understand the physical meaning of $\tmu_{\rm var}(x)$,
we compare it with $\tmu_{\rm BCS}(x) = \mu_{\rm BCS}(x)/t$, the
chemical potential for the {\em unprojected} BCS state
with a gap of $\Delta_{\rm var}$.
$\mu_{\rm BCS}(x)$ is defined via the BCS number equation
$n = 2\sum_{\bk} v_{\bk}^2$, with $n = 1 - x$, 
$\xi_{\bk} = \epsilon(\bk) - \mu_{\rm BCS}$ and 
$E_\bk = \sqrt{\xi_{\bk}^2 + \Delta_{\rm var}^2}$. 
We find, quite remarkably, that except for the immediate vicinity
of $x=0$, over most of the doping
range $\tmu_{\rm var}(x) \approx \tmu_{\rm BCS}(x)$ seen from Fig.~1(b).

\section{Variational phase diagram}

To determine the $T=0$ phase diagram
as a function of doping within our variational approach
we compute: (i) the SC order parameter which allows us to
delineate the SC regime of the phase diagram, (ii) the spin structure
factor which allows us to check for antiferromagnetic long-range order,
and (iii) the low energy optical spectral weight which allows us to
determine whether the system is insulating or conducting. 
Here we describe in detail the 
calculation of the SC order parameter and only mention relevant results
on the spin structure factor and optical spectral weight,
deferring a detailed discussion of the latter to Sec.~X.
We then discuss the three phases -- RVB Mott insulator, d-wave SC and 
Fermi liquid -- and the transitions between them.

\subsection{Superconducting order parameter}

The SC correlation function is the two-particle reduced
density matrix defined by $F_{\alpha,\beta}
(\br-\brp)=\la B^\dag_{\br\alpha} B_{\brp\beta} \ra$,
where the $B^\dag_{\br\alpha}
\equiv \frac{1}{2} (c_{\br\up}^{\dag}c_{\br+\ha\dn}^{\dag}
- c_{\br\dn}^{\dag}c_{\br+\ha\up}^{\dag})$ creates a singlet
on the bond $(\br,\br+\ha)$.
The SC order parameter $\Phi$ is defined in terms of
off-diagonal long-range order (ODLRO) in this correlation:
$F_{\alpha,\beta} \to \pm \Phi^2$ for large $\vert \br - \brp \vert$. 
The $+$ ($-$) sign obtained for
$\ha \parallel$ ($\perp$) to $\hb$, indicating d-wave SC.
In the more familiar fixed-phase representation, 
$\Phi$ would correspond to 
$\vert\la c_{\br\up}^{\dag} c_{\br+\ha\dn}^{\dag}\ra\vert$.

\begin{figure}
\begin{center}
\vskip-2mm
\hspace*{1mm}
\epsfig{file=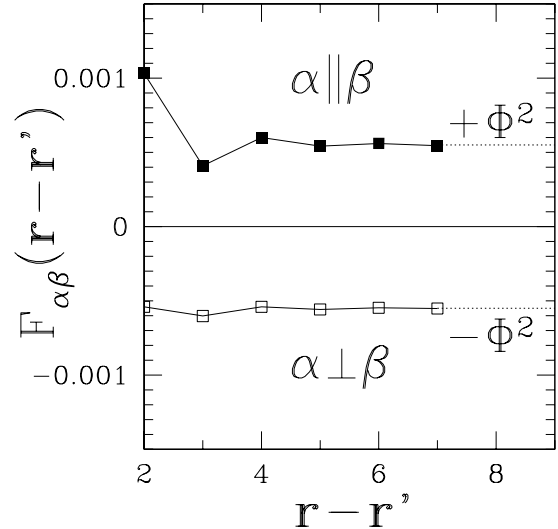,height=3.0in,width=3.0in,angle=0}
\vskip1mm
\caption{Plot of the SC correlation function
$F_{\alpha\beta}(\br - \brp)$, with 
$\ha$ and $\hb$ either ${\hat x}$ or ${\hat y}$ calculated on a 
$15^2+1$ system at $x\approx 0.07$, with $\br-\brp$ along $\hx$.
The correlation function saturates to $\pm \Phi^2$ as indicated
by the dotted line, and defines the d-wave order parameter $\Phi$.}
\label{figphir}
\end{center}
\end{figure}

In Fig.~\ref{figphir} we plot $F_{\alpha\beta}(\br-\brp)$ as a function
of $\vert \br - \brp \vert$ for a hole-doping $x\approx 0.07$. 
For simplicity, we show here the results for $F$ calculated to zeroth order 
in $t/U$, i.e., using $\exp(iS)=1$. We have checked that the much more 
involved calculation which keeps $t/U$ terms leads to only small 
quantitative changes in the results.
We obtain the SC order parameter for various
doping values and plot $\Phi(x)$ in Fig.~\ref{figphase}(a).
In strong contrast to the variational ``gap'' parameter $\tDelta_{\rm var}$,
which was a monotonically decreasing function of $x$ (see 
Fig.~~\ref{figdeltamu}(a)), we find that the order parameter
$\Phi(x)$ is nonmonotonic and vanishes at both 
$x=x_c \approx 0.35$ and at $x=0$.
The vanishing of $\Phi(0)$ was first noted 
by Gros in Ref.~\cite{gros88}. 

\subsection{Phase diagram}

\noindent (1) {\bf Fermi Liquid ($x > x_c$):}
For large doping values $x > x_c \approx 0.35$,
$\tDelta_{\rm var}=0$ implies that there is no pairing and
$\Phi = 0$ implies that there is no superconductivity.
The ground state wavefunction for $x > x_c$ is then 
a Landau Fermi liquid. This can be explicitly checked from its
momentum distribution which shows a sharp Fermi surface with
a finite jump discontinuity all around the Fermi surface.

\begin{figure}
\begin{center}
\vskip-2mm
\hspace*{1mm}
\epsfig{file=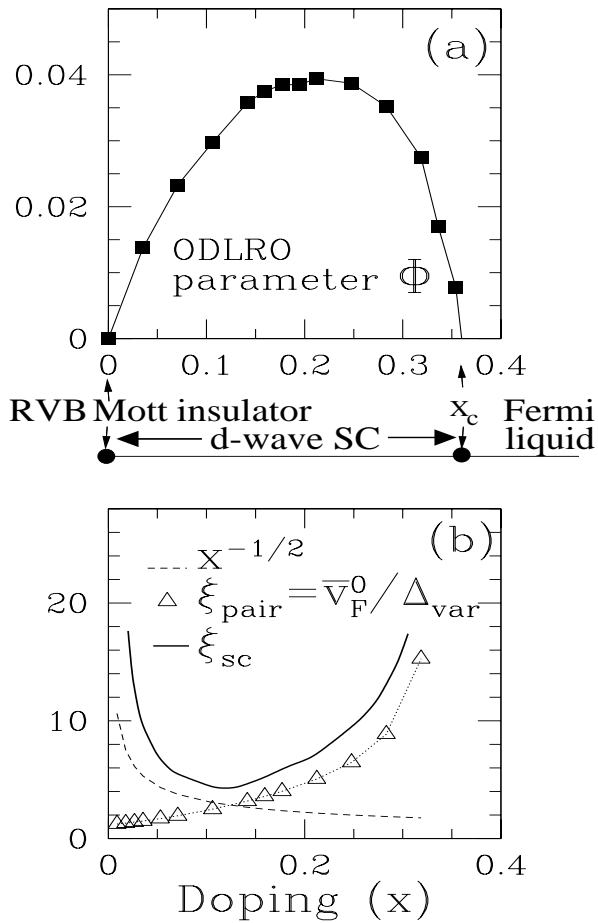,height=5.0in,width=3.1in,angle=0}
\vskip1mm
\caption{Phase diagram obtained within
our variational calculation is shown between panels (a) and (b).
The phases are a spin-liquid Mott
insulator at $x=0$, a correlated d-wave SC for $0 < x < x_c$, 
and a Fermi liquid metal for $x > x_c$.
(a) Doping dependence of the d-wave order parameter $\Phi(x)$
showing a superconducting ``dome'' with optimal doping around $x \simeq 0.2$. 
(b) Doping dependence of various length scales:
the ``pair size'' $\xi_{\rm pair}=\bar{v}^{0}_{_F} /\Delta_{\rm var}$ 
is shown as open triangles; the average interhole separation $x^{-1/2}$
is shown as a dashed line; and the SC coherence length
$\xi_{\rm sc}\geq {\rm min}(x^{-1/2}, \xi_{\rm pair})$.
}
\label{figphase}
\end{center}
\end{figure}

\medskip
\noindent (2) {\bf d-Wave Superconductor ($0 < x < x_c$):}
As $x$ decreases below $x_c$, a non-zero 
$\tDelta_{\rm var}$ indicates that pairing develops and
leads to d-wave superconducting order characterized by $\Phi$.
The most striking result is the qualitative difference
between the doping dependence of the variational
$\tDelta_{\rm var}$ and the SC order parameter $\Phi$.
Although $\tDelta_{\rm var}$ increases monotonically with underdoping
(i.e., decreasing $x$), $\Phi$ reaches a maximum near $x=0.2$ and 
then goes down to zero as $x \to 0$. We shall show later in Sec.~X
that the superfluid stiffness also vanishes as $x \to 0$.

{\em Why} does the SC order parameter $\Phi$
vanish at half-filling even though the pairing amplitude 
$\tDelta_{\rm var}(x=0)$ is non-zero? 
We give two arguments to understand how Mott physics (no-double occupancy) 
leads to the loss of superconductivity as $x \to 0$.
First, projection leads to a fixed electron number $n_{\br}=1$ at each site
when $x=0$, thus implying large fluctuations in the conjugate variable, 
the phase of the order parameter. These quantum phase fluctuations destroy 
SC ODLRO leading to $\Phi(x=0)=0$. 

Quite generally, we expect that the order parameter 
$\Phi(x)$ should be proportional to $\tDelta_{\rm var}(x)$. 
However an additional $x$-dependence arises from projection. 
The correlation function $F_{\alpha,\beta}(\br - \brp)$ 
involves moving a pair of electrons on adjacent sites to a distant pair 
of neighboring sites, which should both be vacant in order to satisfy the 
no-double-occupancy constraint. Since the density
of vacant sites (holes) $\sim x$, the probability to find two 
holes implies $F \sim x^2$, leading to an additional factor
of $\Phi \sim x$. Putting these two effects together we get
$\Phi \sim x \tDelta_{\rm var}(x)$ 
which agrees remarkably well with the calculated
non-monotonic $\Phi(x)$.

The dome in $\Phi(x)$ seen in Fig.~\ref{figphase}(a)
naturally leads to the notion of optimal doping near $x=0.2$
where SC correlations are strongest. Based on our $T=0$ calculation,
we expect that the transition temperature $T_c(x)$ should correspondingly
also exhibit non-monotonic $x$-dependence, with a maximum at optimal doping.
The SC dome is thus determined in our variational calculation 
by loss of pairing on the overdoped side as $\tDelta_{\rm var}$ vanishes
beyond $x_c$ and by the loss of phase coherence (as we will further 
substantiate below in Section X) due to Mott physics at $x=0$. 

We emphasize that we do {\em not} need to invoke any competing order 
parameter at small $x$, to explain the loss of superconductivity at
low doping. The high-energy Mott constraint of no double-occupancy 
forces this on us, and competing orders (such as antiferromagnetism or 
charge order) which may emerge at low energy scales are 
{\em not} the primary cause for the loss of superconductivity at small $x$.

We can also crudely estimate the ``condensation energy'' by calculating the
energy difference between the projected SC ground state and
the non-SC state defined by the projected Fermi gas.
The very {\em definition} of ``the non-SC
ground state'' is fraught with difficulty.
However, we feel that the projected Fermi gas state
is a physically reasonable candidate on (and only on) the
{\em overdoped side}, i.e, for $x \gtrsim 0.2$.

Computing the ground state energy difference between the
projected SC and the projected Fermi gas, we 
find that it is the AFM superexchange term in $\tilde{\cal H}$
which drives the SC condensation energy.
Our preliminary estimate of the condensation energy at optimal
doping ($x= 0.2$) is $22 \pm 4$ K per unit cell. 
Given the crudeness
of the estimate, particularly in the overestimate of the ``normal
state'' energy as discussed below, it is not surprising that this
result is much larger than the experimental value of order $1K$
per ${\rm CuO}_2$ plaquette \cite{loram,momono}.
It should be emphasized that the projected Fermi gas
has no variational parameters at all and therefore
leads to a rather poor energy estimate even for overdoping.
We will discuss details of the condensation energy calculations
elsewhere \cite{paramekanti03}.
The condensation energy relative to the staggered flux state in the 
optimal and underdoped regime has been discussed in Ref.\cite{ivanovEc}.

\begin{figure}
\begin{center}
\vskip-2mm
\hspace*{1mm}
\epsfig{file=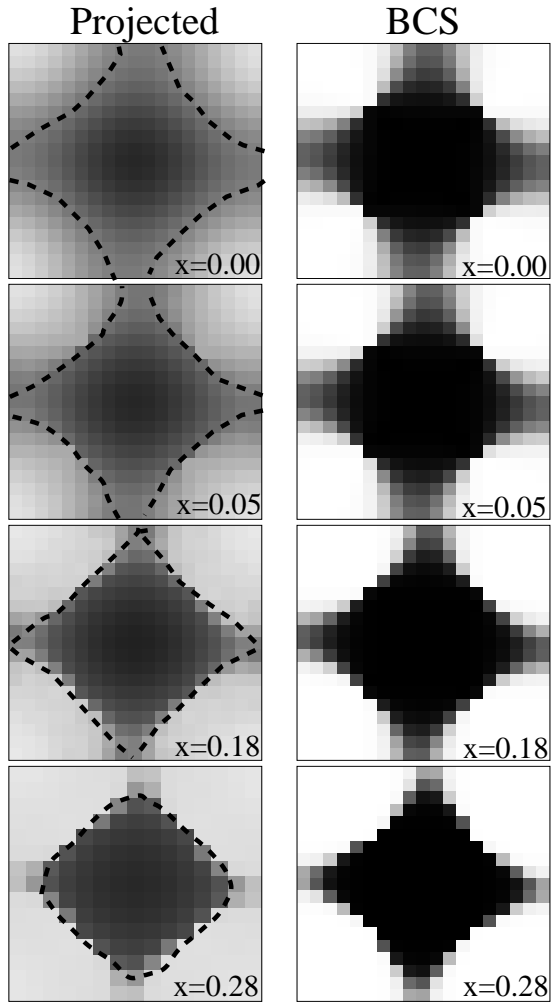,width=3.0in,angle=0}
\vskip1mm
\caption{Grayscale plots (black $\equiv 1$, white $\equiv 0$)
of the momentum distribution $n(\bk)$ at various dopings $x$. 
The left panel shows the results of the projected
variational calculation. The dashed line marks the contour on
which $n(\bk)=1/2$, which closely resembles the non-interacting
Fermi surface. The right panel shows the $n(\bk)$ of the
unprojected BCS calculation at the corresponding doping value.}
\label{figgraynk}
\end{center}
\end{figure}

\bigskip

\noindent (3) {\bf Mott Insulator ($x=0$):}
At half-filling ($x=0$) $\Phi=0$ implying that the undoped state is
nonsuperconducting. We will show below in Section X that its
low frequency integrated optical spectral weight vanishes and thus it is an
insulator. 

A careful finite size scaling analysis
of the spin structure factor shows that the $x=0$ is a critical state 
exhibiting {\em algebraic decay} of antiferromagnetic (AFM) spin correlations:
$\la S^z_\br S^z_{\bf 0}\ra \sim (-1)^{r_x+r_y}/|\br|^{3/2}$.
These results will be presented elsewhere \cite{paramekanti03}
along with a detailed discussions of other competing order parameters
at low doping. 

The variational state at $x=0$ is an insulator made up of a 
superposition of singlet pairs: since $\tDelta_{\rm var}(x=0)$ is nonzero, 
the function $\varphi(\br - \brp)$ describes the singlet bonds in this
state. The Gutzwiller projection prevents this liquid of singlet pairs
from (super)conducting, and the $x=0$ state is an resonating
valence bond (RVB) \cite{anderson87} or spin-liquid Mott insulator. 

The form of the wavefunction studied here apparently 
does not have enough variational freedom to exhibit broken spin-rotational 
or translational invariance to describe the Neel AFM state which is 
known to be the experimental ground state of the undoped cuprate materials and 
also believed to be the ground state of the 2D large $U$
Hubbard model at half-filling. 
We plan to study in the future the competition between 
SC and AFM by adding more variational freedom in our trial state, but 
our primary focus here is a detailed characterization of
the simplest description of a strongly-correlated
{\em superconducting} state.

We should also note that the ground state energy of the our spin-liquid state
at $x=0$ is within few percent of the best estimates. The
best way to present this comparison to look at the 
spin correlation $\gamma \equiv \langle \bS_\br\cdot\bS_\brp\rangle$
between neighboring sites $\br$ and $\brp$ at half-filling. 
For our state we find $\gamma = -0.313 \pm 0.002$.
For comparison, the best estimate for the 2D 
{\it nearest-neighbor} Heisenberg model
is $\gamma = -0.3346 \pm 0.0001$ from Green's function Monte Carlo
calculations \cite{trivedi90}, while a classical Neel state has
$\gamma = - 0.25$.
For the nearest neighbor hopping Hubbard model in the large $U$ limit,
the ground state energy per site 
is $E_0 = 2J(\gamma - 1/4)$ at half-filling. Further neighbor
hopping leads to additive corrections of order $J^\prime/J = (t^\prime/t)^2
= 1/16$ for our choice of parameters.

\subsection{Phase Transitions and Correlation Lengths}

The variational wavefunction Eq.~(\ref{wavefn}) describes
the three phases discussed above, and our approach
also gives interesting information
about the quantum phase transitions between these phases.
We find that there are diverging length scales in the SC state
as one approaches the Mott insulator at $x=0$ and also the Fermi liquid metal
beyond $x_c$. 

The internal pair wavefunction $\varphi(\bk) = v_\bk/u_\bk$, 
or more correctly the related quantity $v_\bk u_\bk$,
defines a ``pair-size''
$\xi_{\rm pair} = \bar{v}^0_F/\Delta_{\rm var}$,
where $\bar{v}^0_F$ is the bare average Fermi velocity.
Projection is expected not to affect the pair-size much.
$\xi_{\rm pair}$ diverges at $x_c$ and decreases monotonically
with decreasing hole doping as the pairing becomes progressively
stronger. The pair size remains finite at $x = 0$, where
it defines the range of singlet bonds in the RVB
insulator, which is very short, of the order of the lattice spacing. 
%
%
Note that in converting the dimensionless $\tDelta_{\rm var}$
to an energy (needed to define the pair size),
we need to use the scale of either $t$ or $J$ (which we have chosen to be
300 and 100 meV respectively).
In Section IX we will discuss this question in detail; here
we simply chose $\Delta_{\rm var} = t \tDelta_{\rm var}$, since
in any case we want to get a lower bound on the coherence length.

A second important length scale is the average
inter-hole spacing $1/\sqrt{x}$.
At shorter distances there are no holes, no SC order can develop
and the system effectively looks like the $x=0$ insulator. 
The SC correlation length
$\xi_{\rm sc}$ must necessarily satisfy
$\xi_{\rm sc} \ge \max\left(\xi_{\rm pair},1/\sqrt{x}\right)$.
As shown in Fig.~\ref{figphase}(b), this bound implies that $\xi_{\rm sc}$
must diverge both in the insulating limit $x \to 0$
and the metallic limit $x \to x_c^{-}$,
but could be small (few lattice spacings) near optimal doping.

The divergence of $\xi_{\rm sc}(x)$  as $x\to 0$ could
also be tested in experiments designed to measure the conductivity
$\sigma(\bq,\omega)$ in underdoped SC's, at {\em nonzero momentum} $\bq$.
We expect significant $\bq$-dependence at low $\omega$, with the
conductivity rapidly vanishing for $\bq > \xi_{\rm sc}^{-1}$ as
insulating behavior is recovered.
Another important question related to the proper definition of
the correlation length concerns the vortex core radius as function
of doping $x$. This problem deserves careful study using Gutzwiller
wavefunctions, since the $U(1)$ slave boson-gauge theory approach
predicts that the vortex core size diverges as $1/\sqrt{x}$ as $x \to 
0$ \cite{dhlee}, while an SU(2) approach \cite{palee97} suggests a 
stronger $1/x$ divergence.

It is clear that we must carefully distinguish 
between various ``coherence lengths'', which are the
same in simple BCS theory up to factors of order unity, but
could be very different in strongly correlated SCs. Only the
result of a detailed calculation can reveal which coherence
length is relevant for a particular experiment. 

\begin{figure}
\begin{center}
\vskip-2mm
\hspace*{1mm}
\epsfig{file=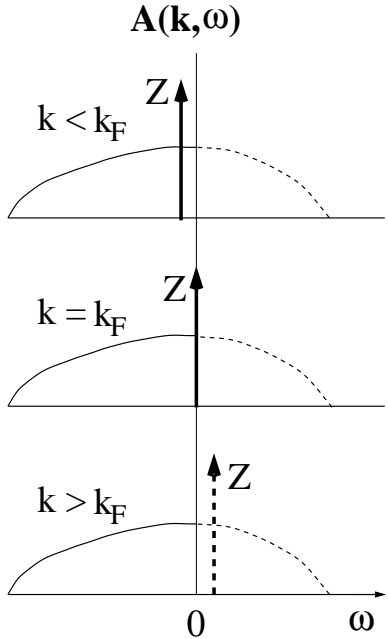,width=2.0in,angle=0}
\vskip1mm
\caption{Schematic plot of the spectral function when gapless
quasiparticles are dispersing across the chemical potential,
$\omega=0$, with a quasiparticle weight $Z$ and velocity $v_{_F}$.
The solid (dashed) lines indicate the occupied (unoccupied) part 
of the spectral function.}
\label{figgaplessQP}
\end{center}
\end{figure}
\section{Momentum Distribution}

Next we study the momentum distribution 
$n(\bk) = \la c^\dag_{\bk\sig} c_{\bk\sig} \ra$. This is calculated
by computing the Fourier transform
of $\la {\cal G}_\sig(\br,\brp) \ra \equiv \la c^\dag_{\br\sig}
c_{\brp\sig} \ra$. The details of the transformed operator 
$\tilde{\cal G} = \exp(iS){\cal G}\exp(-iS)$ 
to first order in $t/U$ are given in Appendix A.

In Fig.~(\ref{figgraynk}) (left panel) we show grayscale plots of $n(\bk)$ at
various doping values ranging from the insulating state at $x=0$ to the
overdoped SC at $x=0.28$. We see that $n(\bk)$ has considerable
structure at all dopings including $x=0$. 
These results are qualitatively consistent with photoemission 
experiments on the SC cuprates \cite{ding97,fujimori00}
and related insulating compounds \cite{ronning98}.

In a strict sense there is no meaning to a
Fermi surface (FS) at $T=0$ since the system is either SC
or insulating (at $x=0$) for $0\leq x < x_c$.
Nevertheless it is interesting to note that if one plots either the contour on 
which $n(\bk)=1/2$ (shown as a dashed line in Fig.~(\ref{figgraynk}))
or the contour on which $\vert\nabla_\bk n(\bk)\vert$ is maximum 
(not shown), both are very similar to the {\it non}interacting FS 
that would have been obtained from the free dispersion $\epsilon(\bk)$. 
For this reason we call such contours the interacting ``Fermi surface''
\cite{mesot01}.
These similarity between the interacting and non-interacting
FS is closely related to the approximate equality of $\mu_{\rm BCS}$ and the 
variational parameter $\mu_{\rm var}$ 
discussed in Sec.~IIIB.

To further see the extent to which strong correlations affect the
momentum distribution, we compare the $n(\bk)$ obtained from the
projected wavefunctions in the left panel with that obtained from
simple BCS theory. The BCS result $n(\bk) = v_\bk^2$ using 
optimal values $\tDelta_{\rm var}$ and $\tmu_{\rm var} \simeq \mu_{\rm BCS}/t$ is
plotted in the right panel of Fig.~(\ref{figgraynk}).
While projection leads to a transfer of spectral weight
(i.e., $n(\bk)$ intensity) from the zone center to the
zone corners, the overall ``topology'' of the momentum distribution
is not qualitatively changed.

For $t^\prime=t/4$, the {\it non}interacting FS and the interacting
``FS'' both show a change in topology from a large hole-like barrel
centered at $(\pi,\pi)$ for small $x$ to an electron-like FS for
$x\gtrsim 0.22$. The precise value at which the topology changes
depends sensitively on the sign and value of $t^\prime$. 
Such a topology change has been clearly observed in ARPES data
on LSCO \cite{fujimori00} and less obviously in BSCCO where the
topology change may be happening at large overdoping \cite{jc_private}.

In later sections we will return to a detailed study of $n(\bk)$ along
special directions in the Brillouin zone, where we will see that
strong correlations play a crucial role, even though they appear to be
not very important in so far as gross features like the topology
of the ``Fermi surface'' is concerned.

\begin{figure}
\begin{center}
\vskip-2mm
\hspace*{1mm}
\epsfig{file=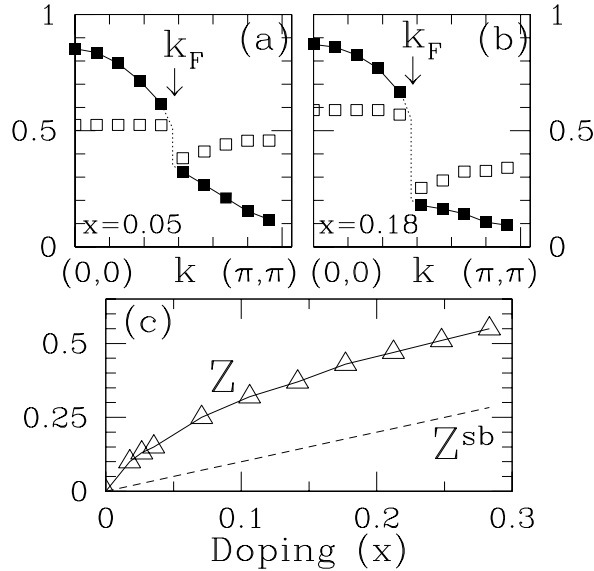,height=3.0in,width=3.1in,angle=0}
\vskip1mm
\caption{(a),(b) The momentum distribution, $n(\bk)$, along the nodal
direction $(0,0)\to(\pi,\pi)$ (black squares). The white squares are
results for the tJ model, and correspond to ignoring $t/U$ corrections
in Eq.~(\ref{nk1}). The nonmonotonic behavior of $n(\bk)$ near
$k_{_F}$ is removed on including ${\cal O}(t/U)$ terms.
The discontinuity in $n(\bk)$ at $k = k_{_F}$ 
signals gapless quasiparticles with a weight $Z$ 
determined by the magnitude of the jump.
(c) Doping dependence of $Z$ compared with $Z^{\rm sb}=x$
from slave-boson mean-field theory. }  
\label{figdiagnk}
\end{center}
\end{figure}
\section{Spectral Function Moments}

A variational wave function approach is limited to the calculation of 
equal-time correlations and thus interesting dynamical information would,
at first sight, seem to be out of reach. 
We now show that this is not always true.
First, the frequency moments of dynamical correlations can always be written 
as equal-time correlators, and this in itself can give very useful information 
as we shall see in the following sections. Further, 
one can obtain much more detailed information, when
the moments, which are functions only of $\bk$ with $\omega$ integrated
out, exhibit singularities in $\bk$. In the case of the single-particle
Green's function, we show that the singularities of its moments at $T=0$
are completely governed by gapless quasiparticles, if they exist.

The one-particle spectral function is defined in terms of the 
retarded Green function as $A(\bk,\omega) = - {\rm Im}
G (\bk,\omega + i 0^+)/\pi$, and has the $T=0$ 
spectral representation
\bea
A(\bk,\omega) & =&  \sum_m\big[ \vert\la m\vert c_{\bk\sig}^\dag\vert
0\ra\vert^2 \delta(\omega+\omega_0-\omega_m) \nn \\
&+& \vert\la m\vert c_{\bk\sig}\vert
0\ra\vert^2 \delta(\omega-\omega_0+\omega_m) \big].
\eea
Here $\vert m \ra$ (and $\omega_m$) are the eigenstates (and eigenvalues) of 
$({\cal H} - \mu N)$ with $N$ the total number of particles and all 
energies are measured with respect to $\mu$. 

One can consider moments of the full spectral function \cite{white91},
but for our purposes it is much more useful to consider moments of the
{\em occupied part} of spectral function $f(\omega) A(\bk,\omega)$, 
where $f(\omega)$ is the Fermi function. This is also the quantity 
measured in ARPES experiments \cite{randeria95}. At $T=0$, 
$f(\omega) = \Theta(-\omega)$ and only the second term in the spectral
representation contributes:
$\Theta(-\omega)A(\bk,\omega) = \vert\la m\vert c_{\bk\sig}\vert 0\ra\vert^2 
\delta(\omega-\omega_0+\omega_m)$.

The $\ell^{\rm th}$ moment
of the occupied spectral function 
$M_\ell(\bk)\equiv \int_{-\infty}^0 d\omega \omega^\ell A(\bk,\omega)$ 
can be expressed as a ground state correlator following standard algebra.
We will focus on the first two moments in what follows.
These are given by:
\bea
M_0(\bk) = \int_{-\infty}^{0}d\omega A(\bk,\omega)& =&
\sum_m \vert\la m \vert
c_{\bk\sig} \vert 0\ra\vert^2 = n(\bk) \nn \\
M_1(\bk) = \int_{-\infty}^{0}d\omega \omega
A(\bk,\omega)& = &\sum_m (\omega_0 - \omega_m) \vert\la m \vert
c_{\bk\sig} \vert 0\ra\vert^2 \nn \\
&=& \la c_{\bk\sig}^\dag
\big[ c_{\bk\sig}, {\cal H-\mu N}\big] \ra \nn \\
&=& \la c_{\bk\sig}^\dag
\big[ c_{\bk\sig}, {\cal H} \big]\ra - \mu n(\bk)
\label{M1}
\eea

We next describe the characteristic singularities in these
moments arising from coherent quasiparticle (QP) excitations;
the result for the momentum distribution is very well known,
but that for the first moment seems not to have been appreciated before.
In the presence of gapless quasiparticles, the spectral function has
the form
\be
A (\bk,\omega)
= Z \delta(\omega-\tilde{\xi}_\bk) + A_{\rm inc}(\bk,\omega),
\ee
plotted schematically in Fig.~(\ref{figgaplessQP}). 
Here $Z$ is the coherent QP weight  $(0<Z\le1)$ and 
$\tilde{\xi}_{\bk}=v_{_F} (k-k_{_F})$ is the QP dispersion
with $k_{_F}$ the Fermi wavevector and $v_{_F}$ the Fermi velocity.
$A_{\rm inc} (\bk,\omega)$ is the
smooth, incoherent part of the spectral function.
It is then easy to see from Eq.~(\ref{M1}) that
\bea
M_0(\bk) &=& n(\bk) = Z \theta(-\tilde{\xi}_\bk) + \ldots\nn \\
M_1(\bk) &=& Z \tilde{\xi}_\bk \theta(-\tilde{\xi}_\bk) + \ldots,
\eea
where the terms omitted are the non-singular contributions
from the incoherent piece.

It follows that, precisely at $k=k_{_F}$,
$M_0(\bk)$ has a jump discontinuity of $Z$
and $d M_1(\bk)/d k$ has a discontinuity of $Z v_{_F}$.
Thus, studying the moments of
$\Theta(-\omega)A(\bk,\omega)$ allows us to extract
$Z$ and $v_{_F}$ from singular behavior of
$M_0(\bk)$ and $M_1(\bk)$, while $k_{_F}$ can be determined from the
location in $\bk$-space where this singularity occurs.

It is worth emphasizing what has been achieved.
In strongly correlated systems, interactions lead to a transfer of
spectral weight from coherent excitations to incoherent features
in the spectral function. The values of $M_\ell(\bk)$
are, in general, dominated by these incoherent features (which
we will find to be very broad in the cases we examine), but
nevertheless their singularities are governed by the gapless coherent part 
of the spectral function, if it exists. We exploit
these results below in our study of nodal quasiparticles in the d-wave
SC state.

\begin{figure}
\begin{center}
\vskip-2mm
\hspace*{1mm}
\epsfig{file=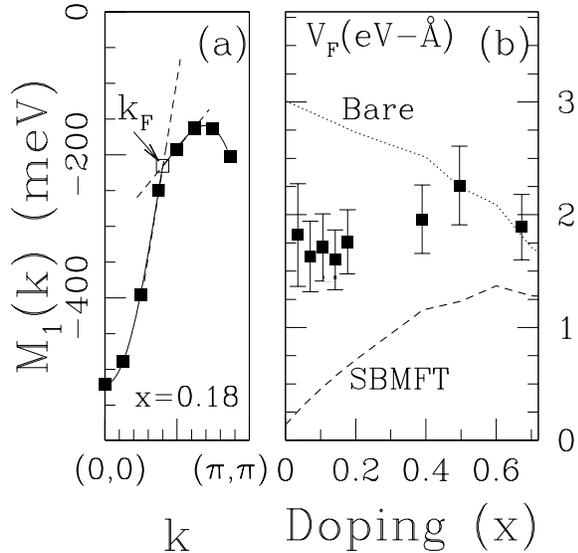,height=3.2in,width=3.1in,angle=0}
\vskip1mm
\caption{(a) The first moment of the occupied part of the spectral function 
$M_1(\bk)$ along the zone diagonal $(0,0)\to (\pi,\pi)$ 
showing discontinuity of $Z v_{_F}$ in its slope, 
$d M_1(\bk)/dk$, at $k=k_{_F}$. (b) Doping dependence of $v_{_F}$, extracted
from $M_1(\bk)$. The error bars here are associated
with fits to $M_1(\bk)$ and $n(\bk)$ and errors in $Z$. Also shown are the
bare velocity $v^0_{_F}$ (dashed line) and the
QP velocity within slave-boson mean-field theory, $v_{_F}^{\rm sb}$
(dotted line). The experimental QP velocity $\simeq 1.5 eV\AA$ from
ARPES data\protect\cite{kaminski} in near optimal BSCCO, and
experiments indicate that it is nearly doping independent
.\cite{nodalQPpaper}}
\label{fignodal}
\end{center}
\end{figure}
\section{Nodal quasiparticles}

There is considerable evidence from ARPES 
\cite{kaminski,valla99,bogdanov00}
and transport experiments \cite{hardy93,ong95,taillefer00} that
there are sharp gapless quasiparticle (QP) excitations in the low
temperature superconducting state along the nodal direction $(0,0)\to
(\pi,\pi)$. These nodal excitations then govern the low temperature
properties in the SC state.
In this section we show that our SC wavefunction supports sharp nodal
quasiparticles and calculate various properties such as their
location $k_{_F}(x)$, spectral weight $Z(x)$ and Fermi velocity
$v_{_F}(x)$ as a function of doping and compare with existing experiments.

\subsection{$k_{_F}(x)$ and $Z(x)$ }

In Figs.~\ref{figdiagnk}(a) and (b) we plot the momentum distribution
(black squares) $n(\bk)$ along $(0,0) \to (\pi,\pi)$ 
for two different doping levels. We see a clear jump discontinuity 
which implies the existence of sharp, gapless nodal QPs.
(Note that such a discontinuity is 
{\it not} observed along any other direction in $\bk$-space
due to the existence of a d-wave SC gap.) 
We thus determine the nodal $k_F(x)$, the location of the discontinuity
in $n(\bk)$, and the nodal QP weight $Z$ from the magnitude
of the jump in $n(\bk)$ \cite{footnote_z}.

We find that the nodal $k_F(x)$ has weak doping dependence consistent 
with ARPES \cite{ding97,fujimori00}, and
at optimal doping $k_{_F}\approx 0.69 \AA^{-1}$, which is close to
the ARPES value of $0.707 \AA^{-1}$ \cite{mesot99}.
As already noted while discussing Fig.~(\ref{figgraynk}),
$k_{_F}$ is not much affected by interaction and 
noninteracting (band theory) estimates for $k_{_F}$ are accurate.

In contrast, we find that interactions have a very strong effect on
coherence: the QP spectral weight $Z$ is considerably reduced from
unity and the incoherent weight $(1-Z)$ is spread out to high energies.
We infer large incoherent linewidths from the fact that, even 
at the zone center $\bk = (0,0)$ which is the ``bottom of the band'' 
$n(\bk = (0,0)) \simeq 0.85$ (for $x = 0.05$),
implying that 15\% of the spectral weight must have
spilled over to the unoccupied side $\omega > 0$. 
A second indicator of large linewidths is the magnitude of the
first moment discussed below.

The doping dependence of the nodal QP weight $Z(x)$ is shown in 
Fig.~\ref{figdiagnk}(c). The most striking feature is the complete
loss of coherence as $x \to 0$, with $Z\sim x$ as the 
insulator is approached.  We can understand the vanishing
of $Z(x)$ as $x \to 0$ from the following argument. 
A jump discontinuity in $n(\bk)$ leads to the following
long distance behavior in its Fourier transform
${\cal G}(\br) = \la c^\dagger_\sigma(\br) c_\sigma(0)\ra$:
a power law decay with period $k_{_F}^{-1}$ oscillations 
and an overall amplitude of $Z$. 
However, for ${\cal G}(\br)$ to be nonzero at {\it large} $r$ 
in a projected wavefunction, we need to find a vacant site at a point 
$\vert \br \vert$ away from the origin. This probability scales as $x$, 
the hole density, and thus $Z \sim x$. 
Near half-filling we expect this $x$-dependence due to projection to dominate
other sources of $x$-dependence \cite{footnote_log}, in the same way
as the discussion of the order parameter $\Phi(x)$ (see Sec.~V.B)
and we see why $Z \sim x$ for $x \ll 1$.

After our theoretical prediction \cite{paramekanti01} of the nodal QP $Z(x)$,
ARPES studies on La$_{2-x}$Sr$_x$CuO$_4$ (LSCO) 
\cite{yoshida02} have been used to systematically extract 
the nodal $Z$ as a function of Sr concentration $x$. 
The extracted $Z$'s are in arbitrary units,
but the overall trend, and particularly the vanishing of $Z$ as
$x \to 0$, is roughly consistent with our predictions. We should
note however that underdoped LSCO likely has a strong influence of
charge and spin ordering competing with the superconductivity.
While this physics is not explicitly built into our 
wavefunction, the vanishing of $Z$ with underdoping is a general
property of projected states as discussed above.

We next contrast our results with those obtained for similar
variational calculation on the tJ model, which gives insight into
the differences between the large $U$ Hubbard (black squares)
and tJ models (open symbols in Figs.~\ref{figdiagnk}(a) and (b)).
The tJ model results, which set $\exp(iS)=1$ in so far as the operator
$c^\dagger_\sigma(\br) c_\sigma(0)$ is concerned, lead to an 
$n(\bk)$ which is a non-monotonic function of $\bk$. 
This is somewhat unusual, although not forbidden by any exact inequality 
or sum rule. Further, the tJ model $n(\bk)$ is a $\bk$-independent
constant equal to one-half at $x=0$.
However, we find that the $t/U$ corrections incorporated in
the $\exp(iS)$ factor in the Hubbard model, which arise from mixing in
states with double occupancy into the ground state, lead to a
nontrivial structure in $n(\bk)$ at all $x$ including $x=0$
and also eliminate the nonmonotonic feature near $k_{_F}$.
At $x=0$ (see Eq.~(\ref{nkspin}) in Appendix B)
the $t/U$ corrections arise from short-range antiferromagnetic spin
correlations. These correlations are expected to persist even away from
half-filling although they get weaker with increasing $x$.

Finally we compare our result for $Z(x)$ with that obtained from
slave boson mean field theory (SBMFT).
In Fig.~\ref{figdiagnk}(c), we also plot the SBMFT result $Z^{\rm sb}=x$ 
and find $Z(x) > Z^{\rm sb}(x)$, i.e., the SBMFT underestimates
the coherence of nodal QPs. To understand the significance of this
difference, we must look at the assumptions of SBMFT.
(1) There is a full ``spin-charge separation'', so that
that the spinon and holon correlators {\it completely} decouple, i.e.,
factorize.
(2) There is {\it complete} Bose condensation of the holons: 
$\vert\la b \ra\vert^2=x$.
(3) The spinon momentum distribution corresponds to that of a {\it mean-field}
d-wave BCS SC with a jump $Z_{\rm sp}=1$ along the nodal direction.
With these assumptions, the jump in $n(\bk)$ is just given by
$Z^{\rm sb} = \vert\la b \ra\vert^2 Z_{\rm sp} = x$.
Conditions (2) and (3) are definitely
violated as one goes beyond the mean field approximation.
However, we then expect $\vert\la b \ra\vert^2 < x$
and $Z_{\rm sp} < 1$ both of which lead to a 
{\it further decrease} in $Z^{\rm sb}$.
Thus the observed inequality $Z(x) > Z^{\rm sb}(x)$ implies that
assumption (1) must also be violated when the constraint is taken into
account by including gauge fluctuations around the SBMFT saddle point. 
Inclusion of such gauge fluctuations would also be necessary to obtain the 
incoherent part of the spectral function.
We thus conclude that the spinons and holons of SBMFT must be 
strongly interacting and their correlator cannot be factorized.

\subsection{Nodal quasiparticle velocity $v_{_F}(x)$}

The first moment $M_1(\bk)$ of the
occupied part of $A(\bk,\omega)$ is plotted
in Fig.~\ref{fignodal}(a) as a function of $\bk$ long
the zone diagonal $(0,0) \to (\pi,\pi)$. 
We note that, even {\it at} $k_{_F}$, the moment
$M_1(k_{_F})$ lies significantly below $\omega = 0$: for $x=0.18$
it is $200$ meV below the chemical
potential. This directly quantifies the large incoherent 
linewidth alluded to earlier.

We have already established the existence of nodal quasiparticles,
so they must lead to {\em singular behavior} in $M_1(\bk)$ 
with a slope discontinuity of $Z v_F$ at $k_F$.
We can see this clearly in Fig.~\ref{fignodal}(a) and
use this to estimate the nodal Fermi velocity $v_{_F}$,
whose doping dependence is plotted in Fig.~\ref{fignodal}(b). 
The large error bars on the $v_{_F}$ estimate come from
the errors involved in extracting the slope discontinuity
in $M_1(\bk)$.

First, we see that $v_{_F}(x)$ is reduced 
by almost a factor of two relative to its bare (band structure)
value $v^0_{_F}$ in the low doping regime of interest, 
which corresponds to a mass-enhancement due to interactions.
At large $x \ge 0.5$, deep in the Fermi liquid regime,
the $v_{_F}$ obtained from the moment calculation agrees with the bare
velocity, which also serves as a nontrivial check on our calculation.

More remarkably, we see that the renormalized $v_{_F}(x)$
is essentially doping independent in the SC part of the phase diagram,
and appears to remain finite as $x \to 0$. Thus, as one approaches the 
insulator at $x=0$, the coherent QP weight vanishes like $Z \sim x$,
but the effective mass $m^\ast$ remains finite. 
This has important implications for the form of the nodal quasiparticle
self-energy which are discussed in detail below.
The value and the weak doping dependence of the nodal Fermi velocity
are both consistent with the ARPES estimate\cite{kaminski} of
$v_F \approx 1.5 eV$-$\AA$ in Bi$_2$Sr$_2$CaCu$_2$O$_{8+\delta}$ (BSCCO).
Very recently, our prediction has been tested by ARPES experiments 
on LSCO \cite{zhou03}, where a remarkably doping independent 
(low energy) $v_F$ has been found.

It is also instructive to compare this result with the 
the SBMFT result $v^{\rm sb}_{_F}$ (dashed line in Fig.~\ref{fignodal}(b))
obtained from the spinon dispersion as discussed in Appendix C.
We find that $v^{\rm sb}_{_F}$ is much less than $v_{_F}$ and 
has considerable doping dependence, even though SBMFT does predict a 
nonzero $v^{\rm sb}_{_F}$ as $x \to 0$. Thus not only do the nodal QPs
have more coherence than in SBMFT, they also propagate faster.

Another important nodal quasiparticle parameter is 
the ``gap velocity'' $v_{2}\equiv
{1 \over k_F} \partial\Delta(\theta)/\partial\theta\vert_{\theta=\pi/4}$, 
which is the slope of the SC energy gap at the node ($\theta=\pi/4$ in the
first quadrant of the Brillouin zone).
Together with $k_{_F}$ and $v_{_F}$, $v_2$ completely specifies the
Dirac cone for the nodal QP dispersion:
$E(\bk) = \sqrt{(v_{_F}k_\perp)^2 + (v_2 k_\parallel)^2}$,
where $k_\perp$ ($k_\parallel$) are the deviations from $k_{_F}$
perpendicular (parallel) to the Fermi surface.
In a d-wave SC one expects the singular part of
$A(\bk,\omega) \Theta(-\omega)$ to be of the form 
$Z v_\bk^2 \delta(\omega + E(\bk))$ near the node.
Thus the singular part of the first moment $M_1(\bk)$ is
given by $-Z\left[ E(\bk)-v_{_F}k_\perp \right]/2$.
For $k_\parallel = 0$ (i.e., $\bk$ along the zone diagonal) this
simply reproduces the slope discontinuity analyzed above.
However, setting $k_\perp = 0$, one finds
$M_1(\bk) = - Z v_2 \vert k_\parallel \vert/2 + {\rm smooth}$,
so that crossing the node by moving along the Fermi surface, one would see
a slope discontinuity in $M_1(\bk)$ from which $v_2$ may
be estimated, in principle.

In practice, we are unable to extract this singularity
from our present calculations owing to two difficulties. 
First, one requires a dense sampling of
$\bk$-points lying on the Fermi surface, which cannot be achieved for 
accessible system sizes which are limited by the computational time.
Second, it is known from experiments \cite{mesot99,taillefer00}
that $v_{_F}/v_2 \gg 1$ (around $15$ to $20$); thus small errors in 
locating the Fermi surface would mean that the $M_1(\bk)$ would be 
dominated by effects of $v_{_F}$. Nevertheless, it would be very 
interesting to calculate $v_2$ in the future.

\subsection{Nodal quasiparticle self-energy}

The doping dependence of nodal QP spectral weight $Z(x)$, and 
Fermi velocity $v_{_F}(x)$ obtained above, places strong constraints 
on the self-energy $\Sigma(\bk,\omega)$ particularly near the
SC to insulator transition as $x\to 0$.
For $\bk$ along the zone diagonal $(0,0)\to(\pi,\pi)$ the
gap vanishes in a d-wave SC and
ignoring the off-diagonal self-energy we can
simply write the Green function as 
$G^{-1}(\bk,\omega)=\omega-\epsilon(\bk)-\mu-\Sigma(\bk,\omega)$, where
$\Sigma \equiv \Sigma^\prime+\Sigma^{\prime\prime}$.
Standard arguments then lead to the results
\bea
Z&=&\left( 1 - \frac{\partial \Sigma^\prime}{\partial\omega} \right)^{-1} 
\nn \\
v_{_F}&=& Z
\left( v_{_F}^0 +\frac{\partial\Sigma^\prime} {\partial k}\right)
\label{selfenergy}
\eea
where the right hand side is evaluated at the node $(k_{_F},\omega=0)$.

From $Z \sim x \to 0$ we conclude that
$|\partial\Sigma^\prime/\partial\omega|$
diverges like $1/x$ as $x\to 0$. 
However since $v_{_F}$ remains finite in this limit, there must be a 
compensating divergence in the $\bk$-dependence of the self-energy with
$\partial\Sigma^\prime/\partial k \sim 1/x$.
A similar situation is also realized in the slave-boson mean-field 
solution discussed in Appendix C, even though it is quantitatively a
poor description of the results for $Z(x)$ and $v_{_F}(x)$.
The first example we are aware of where such compensating divergences 
appeared is the normal Fermi liquid to insulator transition in the large-$N$ 
solution of the tJ model \cite{kotliar.largeN}. 

Note that the results obtained here are very different from many other
situations where the self energy has non-trivial $\omega$ dependence,
but is essentially $\bk$-independent. These include examples as diverse
as electron-phonon interaction, heavy fermions \cite{heavyfermion}
(where the large $m^\ast$ or small $v_{_F}$ is tied to a small $Z$), 
and the Mott transition in dynamical mean field theory \cite{kotliar96}.

\begin{figure}
\begin{center}
\vskip-2mm
\hspace*{1mm}
\epsfig{file=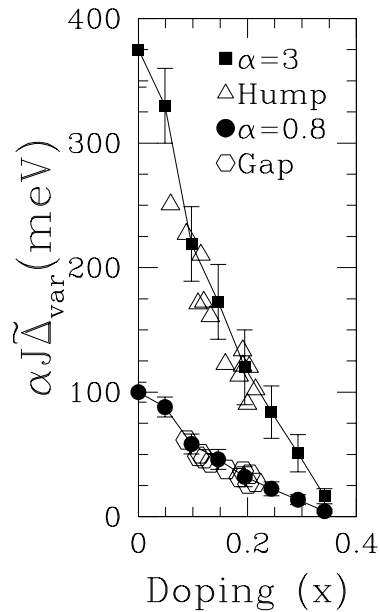,width=2.0in,angle=0}
\vskip1mm
\caption{
The dimensionless variational parameter $\Delta_{\rm var}(x)$ 
of Fig.~1(a) is converted to an energy scale $\alpha J \Delta_{\rm var}(x)$, where
$\alpha$ is a dimensionless prefactor of order unity. The corresponding energy scale
(in meV) is plotted as a function of hole-doping $x$. For the 
choice $\alpha = 0.8$
the energy scale (filled circles) agrees well with the ARPES energy gap in the 
SC state
(open hexagons), while for $\alpha = 3$ it (filled squares) agrees well with 
the ``hump'' scale 
(open triangles) in ARPES spectra at $\bk = (\pi,0)$. All the ARPES results are taken
from Campuzano et al. \cite{campuzano99}.
}
\label{fig_gap_hump}
\end{center}
\end{figure}

\section{Spectral function along $(\pi,0)\to(\pi,\pi)$}


We now move away from the zone diagonal and examine the neighborhood of
$\bk = (\pi,0)$, where the anisotropic d-wave SC gap is the largest.
Our main aim is to see if we can learn something the spectral gap
and its doping dependence. The information available from ground state
correlation functions is not sufficient to rigorously estimate
the excitation gap, so we proceed in two different ways using some guidance from
experiments. First, we convert our dimensionless variational parameter
$\tDelta(x)_{\rm var}$ to an energy scale and compare with experiment, and second,
we use the information available from the moments $n(\bk)$ and $M_1(\bk)$
along $(\pi,0)$ to $(\pi,\pi)$.

As discussed above, the superexchange interaction scale $J$
leads to pairing, and in addition as one approaches the insulator
at $x=0$, $J$ is the only scale in the problem. In view of this it is
natural to consider $J \tDelta_{\rm var}(x)$ as the energy scale characteristic
of pairing.  In Fig.~\ref{fig_gap_hump} we plot 
$\alpha J \tDelta_{\rm var}(x)$ as function of the
hole doping $x$ where $\alpha$ is a dimensionless number of order unity.
For the choice $\alpha = 0.8$ we find that we get very good agreement with the
experimentally measured values and doping dependence of the energy gap
as obtained from ARPES \cite{campuzano99}. 

\begin{figure}
\begin{center}
\vskip-2mm
\hspace*{1mm}
\epsfig{file=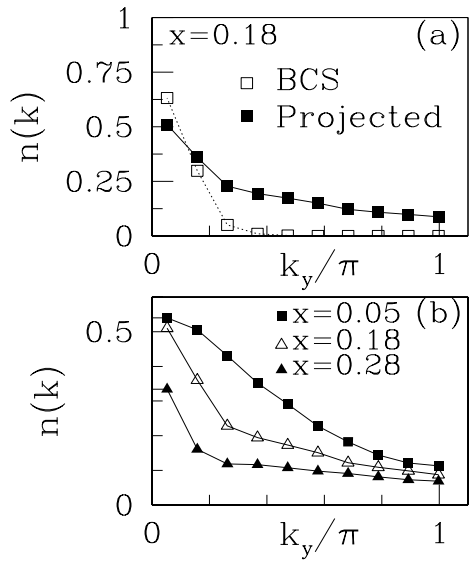,height=3.0in,width=3.1in,angle=0}
\vskip1mm
\caption{(a) The momentum distribution
$n(\bk)$ along the $(\pi,0)\to (\pi,\pi)$ direction,
compared with the unprojected BCS result at the same $\tDelta_{\rm var}$ and
$\mu_{\rm var}$. These results imply that correlations leads to considerable
broadening of $n(\bk)$.
(b) $n(\bk)$ plotted for various $x$, showing increasing
broadening as $x\to 0$, induced by correlations.
}
\label{figpi0.1}
\end{center}
\end{figure}

Next we use the spectral function moments $n(\bk)$ and $M_1(\bk)$ 
to get further insights into the pairing scale.
In the presence of a gap there are no singularities in the moments 
and hence we cannot directly hope to get information about the 
coherent part of the spectral function, as we did near the nodes.
However, as argued below, we will use the moments to determine a 
characteristic energy scale for the incoherent part of the spectral function, 
which we are able to relate, on the one hand, to the variational parameter 
$\tDelta_{\rm var}(x)$ and, on the other, to the
experimentally observed $(\pi,0)$ hump scale in ARPES \cite{campuzano99}.

The ratio of the first moment to the zeroth moment of the occupied 
spectral function
\be
\vert M_1(\bk)\vert/n(\bk) = 
{{\int d\omega \omega f(\omega) A(\bk,\omega)}
\over{\int d\omega f(\omega) A(\bk,\omega)}} \equiv \la \omega \ra(\bk),
\label{ratio_of_moms}
\ee
naturally defines a characteristic energy scale $\omega \ra(\bk)$.
Before studying this quantity in detail, it may help to first look at
each of the moments individually as a function of doping.

From Fig.~\ref{figpi0.1}(a) we see that the
momentum distribution $n(\bk)$ for the projected ground state
is much broader along $(\pi,0) \to (\pi,\pi)$ compared with
that for the unprojected $\vert \Psi_{\rm BCS}\rangle$
with the same $\tDelta_{\rm var}$. This suggests that it is not the energy gap,
but rather the correlation-induced incoherence in the spectral functions,
that is broadening $n(\bk)$. 
A direct measure of the incoherent linewidth in terms of the first moment
$M_1(\bk)$ will be discussed below.
We see that projection leads to a significant build up of
spectral weight for $\bk$'s in the range $(\pi,0.2\pi)$ to $(\pi,\pi)$,
which were essentially unoccupied in the unprojected
$\vert \Psi_{\rm BCS}\rangle$ state.
Correspondingly, correlations lead to a loss of spectral weight near 
$(\pi,0)$. 
The doping dependence of $n(\bk)$ for the projected ground state
is shown in Fig.~\ref{figpi0.1}(b). The increasing importance of
correlations with underdoping is evident from the fact
that $n(\bk)$ becomes progressively broader with decreasing $x$.

\begin{figure}
\begin{center}
\vskip-2mm
\hspace*{1mm}
\epsfig{file=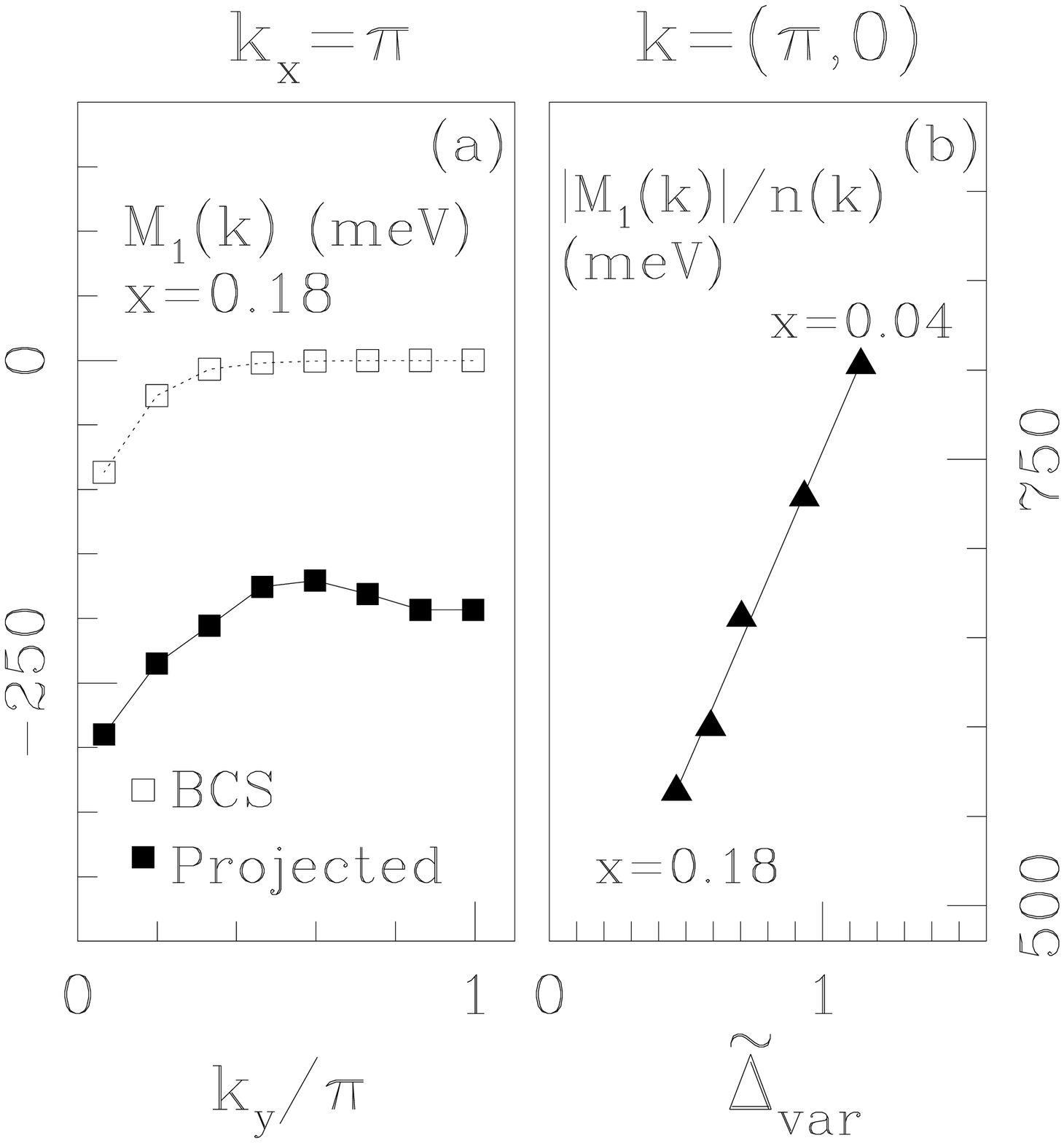,height=3.0in,width=3.1in,angle=0}
\vskip1mm
\caption{
(a) The first moment of the occupied part of the spectral function 
$M_1(\bk)$ along the $(\pi,0)\to (\pi,\pi)$ direction
compared with the corresponding unprojected BCS result.
$\vert M_1(\bk)\vert$ is much larger than
the BCS result from which we infer that strong correlations
lead to very large incoherent linewidth.
(b) Parametric plot of $\la \omega \ra (\bk) \equiv 
\vert M_1(\bk)\vert /n(\bk)$ (see text) at $\bk = (\pi,0)$.
versus $\tDelta_{\rm var}$ with doping $x$ as the implicit parameter.
The linear relation indicates that $\tDelta_{\rm var}(x)$ is
related to an incoherent energy scale in the spectral function
at $(\pi,0)$.}
\label{figpi0.2}
\end{center}
\end{figure}

In Fig.~\ref{figpi0.2}(a), we plot the first moment of the 
occupied part of the spectral function $M_1(\bk)$ along $(\pi,0)\to(\pi,\pi)$
and compare it with the unprojected BCS value. We find
that correlations lead to a large negative value of 
$M_1(\bk)$ which indicates a large incoherent spectral linewidth.

The quantity $\la \omega \ra(\bk)$, defined by the ratio of moments
in eq.~(\ref{ratio_of_moms}), is the characteristic energy scale 
over which the occupied spectral weight is distributed. 
Quite generally we expect this to be dominated by the
large {\em incoherent} part of the spectral function.
We see from Fig.~\ref{figpi0.2}(b) that this energy scale
at $(\pi,0)$ increases with underdoping.
This trend arises from a combination of an 
increasing spectral gap and 
an increasing incoherent linewidth at $\bk=(\pi,0)$ as $x\to 0$. 

It can be argued that the energy $\la \omega \ra(\pi,0)$ 
is an upper bound on the SC gap, even though a very crude (i.e., inaccurate)
one. This can be seen by using $c_{\bk\sigma}\vert\Psi_0\rangle$ as a trial
excited state. A more sensible trial state is obtained by projecting
a Bogoliubov quasiparticle state (where the excitation is created first 
and then projected, unlike the above case where this order is reversed). 
This and further improved excited states are currently under investigation and
will be discussed in a later publication.

From Fig.~\ref{figpi0.2}(b) we see that, as a function of
doping, the energy $\la \omega \ra(\pi,0)$ scales linearly with 
the variational parameter $\tDelta_{\rm var}(x)$, 
characterizing pairing in the wavefunction. 
This suggests that we should think of $\tDelta_{\rm var}$ 
as a characteristic {\it incoherent scale} in the SC state
$A(\bk,\omega)$ at $(\pi,0)$. A second argument in support of 
such an identification comes from the observation that 
at and near $x=0$ $\tDelta_{\rm var}$ is mainly determined by 
minimizing the exchange energy. This implies a close relation between 
{\em local} antiferromagnetic order and short-range d-wave singlet pairing. 
This is directly borne out by correlating the doping dependences of 
$\tDelta_{\rm var}$ and the near-neighbor spin correlation (which will
be described in detail elsewhere).
All of these arguments serve to relate $\tDelta_{\rm var}$ to high energy,
short-distance physics, rather than to the low energy
coherent feature such as the quasiparticle gap in the SC state. 

Motivated by these arguments we compare the energy scale obtained
from the variational parameter
$\alpha J \tDelta_{\rm var}(x) $ with an experimentally observed incoherent
scale in the SC spectral function at $(\pi,0)$. The natural candidate
for the latter is the $(\pi,0)$ ``hump'' in ARPES, where it has been
established that the spectral function at $(\pi,0)$ has a very
interesting peak-dip-hump structure for $T \ll T_c$ 
at all dopings \cite{campuzano99}.
The sharp peak corresponds to the coherent quasiparticle at the
SC energy gap, while ``hump'' comes from the incoherent part of the
spectral function. Two other significant experimental facts 
about the hump are that: 
(a) While both the hump and gap energies decrease monotonically 
with hole doping $x$, their ratio is roughly doping independent, 
with the hump being a factor of
$3.5$ to $4.0$ larger than the SC gap \cite{campuzano99}. 
(b) A vestige of the hump persists even above $T_c$
on the underdoped side where it is called the ``high energy pseudogap''
\cite{campuzano03}. 

In Fig.~\ref{fig_gap_hump}, we find good agreement between 
the energy scale $\alpha J \tDelta_{\rm var}(x)$ with the
ARPES $(\pi,0)$ hump energy measured by Campuzano and coworkers
\cite{campuzano99}, provided we choose\cite{footnote.oldpaper} $\alpha = 3$. 
In summary, with one choice of $\alpha$ ($=0.8$) we find good agreement
with the experimentally observed energy gap and with another choice
of $\alpha$ ($=3.0$) we find good agreement with the ARPES hump 
scale.
We hope that in the future a study of variational {\em excited} states 
will give a direct estimate of the energy gap and also explain the ratio
of approximately $3.75$ (independent of $x$) between the hump and gap.

\begin{figure}
\begin{center}
\vskip-2mm
\hspace*{1mm}
\epsfig{file=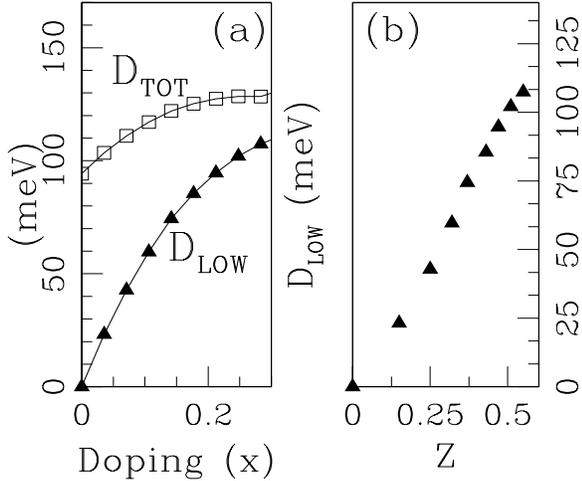,height=3.0in,width=3.1in,angle=0}
\vskip1mm
\caption{(a): Doping dependence of the total ($D_{\rm tot}$) and
Drude or low energy ($D_{\rm low}$) optical spectral weights. 
$D_{\rm low} \sim x$ at low $x$, which implies a Mott insulating state at $x=0$.
(b): Parametric plot of the Drude weight $D_{\rm low}$ versus
the nodal quasiparticle weight $Z$, with hole doping $x$ as the
implicit parameter. We find $D_{\rm low} \sim Z$ over
$0 < x < x_c$.}
\label{figdlow}
\end{center}
\end{figure}

\section{Optical sum rules and superfluidity}

\subsection{Total and low energy optical spectral weights}

We next turn to a discussion of the optical conductivity.
For a superconductor the real part of the optical conductivity
is of the form $\sigma(\omega) = \pi e^2 D_s \delta(\omega)
+ \sig_{\rm reg}(\omega)$, where the condensate contributes
the $\delta(\omega)$ whose strength is the superfluid stiffness $D_s$, 
while the regular part $\sig_{\rm reg}(\omega)$ comes from excitations. 
We will now exploit sum rules which
relate frequency integrals of $\sigma(\omega)$ to equal time
ground state correlation functions which can be reliably calculated in 
our formalism.

For a single-band model, the optical conductivity sum rule 
\cite{scalapino93,paramekanti00} 
can be written as
\be
\int_{0}^\infty d\omega Re \sigma(\omega) =\pi 
\sum_{\bk} m^{-1}(\bk) n(\bk) \equiv \pi D_{\rm tot}/2
\label{dtot}
\ee
where $m^{-1}(\bk) = $ $\left(\partial^2 \epsilon(\bk)/
\partial\bk_x \partial\bk_x \right)$ is the 
{\em noninteracting }inverse mass and we set $\hbar=c=e=1$.
All effects of interactions enter through the momentum 
distribution $n(\bk)$. 

The {\it total} optical spectral weight $D_{\rm tot}(x)$ 
plotted in Fig.~\ref{figdlow}(a) is found to be non-zero
for $x=0$ and an increasing function of hole concentration 
$x$ in the regime shown. We have also found that $D_{\rm tot}$
decreases for $x > 0.4$ and eventually vanishes at $x= 1$,
as it must in the {\em empty band} limit. These results,
which are not shown here, serve as a nontrivial check on our calculation.

It is more important for our present purposes to understand why the
total optical spectral weight in the insulating limit ($x=0$) is
non-zero. This is because the infinite cutoff in the
above integral includes contributions due to transitions
from the ground state to the ``upper Hubbard band'', i.e.,
to states with doubly occupied sites whose energies $\omega \gtrsim U$. 

A physically much more interesting quantity 
is the {\it low frequency} optical weight $D_{\rm low}$, often
called the Drude weight, where the upper cutoff in 
Eq.~(\ref{dtot}) is chosen to be $\Omega_c$ such that
$J \le t \ll \Omega_c \ll U$. The question then arises:
can one write $D_{\rm low}$ as an equal-time correlation?
Toward this end it is convenient to work in the ``low energy'' basis, 
using the ground state wavefunction ${\cal P}\vert\Psi_{\rm BCS}\ra$,
and explicitly include the canonical transformation $\exp(-iS)$
on the operators.  In the presence of a vector
potential, the canonically transformed Hamiltonian (see
Appendix A) to ${\cal O}(t^2/U)$ is given by
\bea
\tilde{{\cal H}}_A&=&\sum_{\br\brp\sig} 
t_{\br\brp}e^{i A_{\br\brp}} \big[h_{\br\sigb} 
c_{\br\sig}^{\dag}c_{\brp\sig} h_{\brp\sigb}\big] \nn \\
&-& \frac{1}{U} \sum_{\br\brp\bR\sig\sigp}
t_{\br\bR} t_{\bR\brp} e^{i \left(A_{\br\bR}+A_{\bR\brp} \right) } 
h_{\br\sigb}
c_{\br\sig}^{\dag} c_{\bR\sig} n_{\bR\sigb} \nn \\
&\times & c_{\bR\sigp}^{\dag} c_{\brp\sigp} h_{\brp\sigpb}.
\label{transformed_H}
\eea
This can be used to extract the diamagnetic response
operator $\tilde{{\cal D}}_{\rm
dia}\equiv\partial^2\tilde{\cal H}_A/ \partial A^2$
\bea
\tilde{{\cal D}}_{\rm dia}&=&\sum_{\br\brp\sig} t_{\br\brp} 
\big[h_{\br\sigb} c_{\br\sig}^{\dag}c_{\brp\sig} h_{\brp\sigb}\big]
(r_x-r'_x)^2 \nn \\
&+& \sum_{\br\brp\bR\sig\sigp}
\frac{t_{\br\bR} t_{\bR\brp}}{U} \big[ h_{\br\sigb}
c_{\br\sig}^{\dag} c_{\bR\sig} n_{\bR\sigb} \nn \\
&\times&
c_{\bR\sigp}^{\dag} c_{\brp\sigp} h_{\brp\sigpb}\big] (r_x-r'_x)^2,
\label{diamagnetic}
\eea
where $r_x$ is the $x$-component of $\br$.
In the low energy projected subspace, standard Kubo formula
analysis shows that the expectation value $\la\Psi_{\rm
BCS}\vert{\cal P} \tilde{{\cal D}}_{\rm dia} {\cal
P}\vert\Psi_{\rm BCS}\ra$ gives both: (i) the diamagnetic
response to a $\bq=0$ vector potential and (ii) the optical
spectral weight in the low energy subspace. 

We thus calculate the low frequency optical spectral weight
\be
D_{\rm low} \equiv {2\over\pi}\int_{0}^{\Omega_c} \!\!\!
d\omega Re \sigma(\omega) =
\la\Psi_{\rm BCS}\vert{\cal P} \tilde{{\cal D}}_{\rm dia} {\cal P}
\vert\Psi_{\rm BCS}\ra
\label{dlow}
\ee
where the last expression is independent of the cutoff
provided $J \le t \ll \Omega_c \ll U$.
$D_{\rm low}$ includes contributions of ${\cal O}(x t)$ from
carrier motion in the lower Hubbard band coming from the first
term in Eq.~(\ref{diamagnetic}), as well as terms
of ${\cal O}(x J)$ from carrier motion which occurs through
virtual transitions to the upper Hubbard band coming from
the second term in Eq.~(\ref{diamagnetic}).
We refer the reader to Ref.~\cite{eskes94} for related discussion. 

$D_{\rm low}(x)$ obtained in this manner
is  plotted in Fig.~\ref{figdlow}(a). In marked contrast
to the total spectral weight, the Drude weight $D_{\rm low}(x)$ 
vanishes as $x \to 0$. 
The vanishing of $D_{\rm low}$ at half-filling 
proves that $\vert \Psi_0 \rangle$ 
describes an insulating ground state at $x=0$. 
Its linear $x$ dependence at small $x$
can be easily understood from the form of
Eq.~(\ref{diamagnetic}) and the no-double-occupancy constraint
(using arguments very similar to the ones used earlier
in understanding the small $x$ behavior of the order parameter
and nodal QP weight).
At low doping, we find that $D_{\rm tot}$ is a weak function of $x$, while
$D_{\rm low}$ increases more rapidly. This reflects a rapid
transfer of spectral weight from the upper to the lower
Hubbard band with increasing hole doping, with a comparatively smaller
change in the total spectral weight.  

There is considerable experimental data on the Drude weight of 
cuprates and its doping dependence; 
see, e.g., Refs.~\cite{orenstein90,cooper93}.
This is usually presented in terms of the plasma frequency 
$\omega_p^*$ defined so that
the integral in Eq.~(\ref{dlow}) is 
$(\omega_p^*)^2/4\pi = D_{\rm low}(e^2/d)K$. Here, $d$ is the
c-axis lattice spacing with $K$ planes per unit cell, and the charge
$e$ and factors of lattice spacing
have been reinstated to convert to real units.\cite{footnote.units}

First, the experimental $(\omega_p^*)^2$ vanishes linearly with the
hole density in the low doping regime, in agreement with our results 
for $D_{\rm low}$.  Furthermore, the data summarized in
Ref.~\cite{cooper93}
gives $\omega_p^* = 2.12 eV$ along the a-axis
(no chains) for YBa$_2$Cu$_3$O$_{6+\delta}$ at optimal doping, i.e. 
$\delta=1$.
Using our calculated $D_{\rm low} \approx 90 meV$ (at optimality), together
with a c-axis lattice spacing $d= 11.68 \AA$ and two planes per unit
cell as appropriate to YBCO, we find $\omega_p^* = 1.67 eV$. Thus, both 
the doping dependence and magnitude of $D_{\rm low}(x)$ 
are in reasonable agreement with optical data on the cuprates.

We predict that the
nodal quasiparticle $Z(x)$ scales with the Drude weight $D_{\rm low}(x)$
over the entire doping range in which the ground state is superconducting.
A parametric plot of these two quantities with $x$ as the implicit parameter,
is shown in Fig.~\ref{figdlow}(b), from which we see that
that $D_{\rm low} \sim Z$ over the entire SC range $0 < x < x_c$.
This is a prediction which can be checked by
comparing optics and ARPES on the cuprates. 
We should note that this scaling must break down at larger $x$, since
as $x \to 1$, $Z$ keeps increases monotonically to unity,
while $D_{\rm low} \to 0$,
since it is bounded above by $D_{\rm tot}$ which vanishes in the
empty lattice limit.

A related scaling has already been noted experimentally.
ARPES experiments \cite{feng00,ding01} have shown that the
quasiparticle weight $Z_A$ at 
the {\em antinodal} point near $\bk=(\pi,0)$ scales as a function of doping
with the superfluid density: $\rho_s \sim Z_A$ for $T \ll T_c$.

Finally, these results also have interesting implications for the SC to
insulator transition as $x \to 0$, 
and caution one against naively interpreting
$D_{\rm low} \sim n_{\rm eff}/m^*$
with $n_{\rm eff}$ related to the size of the Fermi surface.
First, as $x \to 0$, $D_{\rm low}$ indeed vanishes, but
the ``Fermi surface'' always remains large, i.e., 
includes $(1+x)$ holes, as seen in Sec.~VI.
Second the effective mass $m^*$ does not diverge but remains
finite and doping independent as $x \to 0$ (see Sec.~VIII).
Thus one needs to actually calculate the correlation
function defining $D_{\rm low}$ and cannot break it up into a
ratio of individually defined quantities $n_{\rm eff}$ and $m^*$.
A second question arises about the fate of the ``Fermi surface''
as $x \to 0$. Although this contour remains large, the
the coherent QP weight $Z$ vanishes as the insulator is
approached at $x = 0$.
(We have actually shown this only for the $Z$ at the node, 
but expect it to hold everywhere on the FS). 


\subsection{Superfluid stiffness} 

We begin by showing that the Drude weight $D_{\rm low}$ is an
upper bound on the superfluid stiffness $D_s$, and then
use this to compare our results with experiments.
There are many ways to see that $D_s \le D_{\rm low}$ and
we mention three. Different ways of looking at this result
may be helpful because the specific form of $D_{\rm low}$
in Eq.~(\ref{dlow}) is not well known in the literature.

First, we use the Kubo formula for the 
superfluid stiffness $D_s = D_{\rm low}-\Lambda_{\perp}$ 
where $D_{\rm low}$ is the diamagnetic response and 
the paramagnetic response $\Lambda_{\perp}$ is the 
transverse current-current correlator 
evaluated in the ``low energy'' (projected) basis. From its spectral
representation \cite{bound} $\Lambda_{\perp} \geq 0$ which implies 
$D_s \leq D_{\rm low}$. It is important to emphasize \cite{bound} that
in the absence of continuous translational invariance
(either due to periodic lattice and/or impurities)
one cannot in general argue that $\Lambda_{\perp}(T=0)$ vanishes. 

In our second proof, we write the optical conductivity sum rule as
\be
D_{\rm low} = D_s + \int_{0^{+}}^{\Omega_c} d\omega \frac{2}{\pi} 
\sigma_{\rm reg}(\omega),
\ee
with $t,J \ll \Omega_c \ll U$. Since $\sigma_{\rm reg}(\omega) \ge 0$ 
it follows that $D_s \leq D_{\rm low}$. 
Finally, it may be illuminating to see this in yet another way by
applying a phase twist $\Theta$ to the system along the
x-axis, say, which raises the ground state energy by an 
amount $\delta E= D_s \Theta^2/2$.
Following Ref.~\cite{bound} let us make the variational ansatz
\be
|\Psi_\Theta\ra = e^{-i S} {\cal P} \exp{(i \sum_{\br} 
\hat{n}(\br)\theta(\br))} |\Psi_{\rm BCS}\ra
\ee
for the ground state of the system with a phase twist, 
choosing a uniformly winding phase $\theta(\br)$ with
$\theta(\br+L\hat{x})=\theta(\br)+\Theta$.
It is straightforward to show that the energy difference 
between this state and $|\Psi_0\ra$ is 
$D_{\rm low} \Theta^2/2$ with $D_{\rm low}$ given by Eq.~(\ref{dlow}).
We thus arrive at the variational estimate $D_s \leq D_{\rm low}$. 

We now use this bound to extract information relevant to 
experimental data on the $T=0$ superfluid density.
First, the vanishing of $D_{\rm low}$ at small $x$ implies that
we get $D_s \to 0$ as $x\to 0$ which is consistent with 
$\mu$SR experiments \cite{uemura89} in the underdoped regime.
Second, we can rewrite the inequality derived above to
obtain a lower bound on the penetration depth
$\lambda_{\rm L}$ which is related to $D_s$ of a two-dimensional
layer via $\lambda^{-2}_{\rm L} = 4 \pi e^2 D_s/\hbar^2 c^2 d_c$, where
$d_c$ is the mean-interlayer spacing along the c-axis in a layered
compound.  Using $d_c = 7.5 \AA$ appropriate to BSCCO and our
calculated value of $D_{\rm low} \approx 90 meV$ at optimality and 
we find $\lambda_{\rm L}^{}\gtrsim 1350 \AA$. The measured value
in optimally doped BSCCO is $\lambda \simeq 2100 \AA$ 
(Ref.~\cite{sflee96}). 
This agreement is quite satisfactory,
given that the $T=0$ superfluid density is expected to be reduced
by two effects which are not included in our theoretical estimate.
The first is impurities and inhomogeneties \cite{bound}, which are 
certainly present in most underdoped
samples \cite{pan,kapitulnik}, and the second is the effect of long wavelength 
quantum phase 
fluctuations which are estimated \cite{paramekanti00} to lead to a 
10 - 20 \% suppression of the superfluid density. 

\section{Implications for the Finite Temperature Phase Diagram} 

All of our calculations have been done at $T=0$. We now discuss the
implications of our results for the finite temperature phase diagram
of the cuprates, especially on the underdoped side.
We have identified the pairing parameter $\tDelta_{\rm var}(x)$
in our wavefunction with the ``high energy pseudogap'' or the
$(\pi,0)$ ``hump'' feature seen in ARPES experiments; 
see Fig.~1(a) and Sec.~IX. This has the same doping dependence as the 
experimentally observed maximum SC energy gap and the pseudogap temperature
$T^*$ \cite{campuzano99}. On the other hand, the doping dependence 
of the SC order parameter $\Phi(x)$ in Fig.~3 closely resembles
the experimental $T_c (x)$. As discussed in Sec.~V.B, 
strong correlations suppress $\Phi \to 0$ as $x \to 0$.
Further, our results in the previous section imply that the superfluid 
stiffness $D_s$ also vanishes as $x \to 0$.

Thus on the underdoped side the pairing gap will survive in 
the normal state \cite{randeria92} above 
the finite temperature phase transition whose
$T_c$ will be governed by the vanishing of $D_s(T)$ \cite{emery95}.
While this much is definitely
true, a quantitative theoretical calculation of the 
pseudogap region of the cuprate phase diagram will necessarily
involve taking into account additional fluctuating orders which
are likely to exist.

\section{Conclusions}

In this paper we have shown that 
the simplest strongly correlated SC wave function is 
extremely successful in describing the superconducting state properties 
of high Tc cuprates and the evolution of the ground state from a Fermi 
liquid  at large doping,  to a d-wave SC down to the Mott insulator 
at half-filling. The SC dome does not require any competing order, 
but is rather a natural consequence of Mott physics at half-filling. 
The dichotomy of a large pairing energy scale and a small superfluid 
stiffness is also naturally explained in our work and
leads to a pairing induced pseudogap in the underdoped region.

We have also obtained considerable
insight into the doping dependence of various physical observables such
as the chemical potential, coherence length, momentum distribution, 
nodal quasiparticle weight, nodal Fermi velocity, incoherent 
features of ARPES spectral functions, 
optical spectral weight and superfluid density.
We will discuss in a separate paper various competing orders 
-- growth of antiferromagnetic correlations, incipient charge instability,
and singular chiral current correlations -- that
arise in our projected wavefunction in the very low doping regime.


\begin{acknowledgments}
We have benefited from interactions with numerous colleagues
during the course of this work, and would particularly
like to acknowledge discussions with Juan-Carlos Campuzano, 
H. R. Krishnamurthy, Tony Leggett, and Joe Orenstein. 
We are grateful to Phil Anderson for his comments on the manuscript 
and, in particular, for emphasizing to us the importance of
dimensionless variational parameters.
M.R. and N.T. would like to thank the Physics Department of the
University of Illinois at Urbana-Champaign
for hospitality and stimulation during the course of the writing of this
paper. Their work at Illinois was supported through 
DOE grant DEFG02-91ER45439 and DARPA grant N0014-01-1-1062.
A.P. was supported by NSF DMR-9985255 and PHY99-07949 and
the Sloan and Packard foundations. We acknowledge the use of computational
facilities at TIFR including those provided by the D.S.T. Swarnajayanti
Fellowship to M.R.
\end{acknowledgments}

\appendix
\section{The canonical transformation}
\bigskip

In this Appendix we first sketch the construction of
the canonical transformation operator $e^{iS}$ defined in
Sec.~III and then give explicit expressions for various canonically
transformed operators $\tilde{\cal Q} \equiv \exp(iS) {\cal Q} \exp(-iS)$
that are used in the paper.

The Hubbard Hamiltonian (1) may be written as ${\cal H}
= {\cal K}_0 + {\cal K}_{+1} + {\cal K}_{-1} + {\cal H}_{\rm int}$,
where ${\cal K}_n$ have been explicitly defined in Eq.~(\ref{Kndefn}).
In the presence of an external vector potential
$A_{\brp\br}=-A_{\br\brp}$ on the link $(\br\brp)$,
the kinetic energy  terms ${\cal K}_n$ are modified via
$t_{\br\brp} \to t_{\br\brp}\exp(i A_{\br\brp})$.
We consider the unitary transformation 
${\cal H}_A\to \tilde{\cal H}_A =
\exp(iS_A){\cal H}_A\exp(-iS_A)$, where the subscripts on ${\cal H}$ and
$S$ denote the presence of the vector potential. 
We determine $S_A$ perturbatively, order by order in $t/U$,
such that $\tilde{\cal H}_A$ has no matrix elements between different
${\cal D}$-sectors at each order. 

The systematic procedure devised in Ref.~\cite{yoshioka88} 
may be trivially generalized to include the vector potential.
To ${\cal O}(t^2/U^2)$ we find the result
$ S_A = S_A^{[1]}+  S_A^{[2]}$ with
\bea
i S_A^{[1]}& = & \frac{1}{U} \big({\cal K}_{A,+1} - {\cal K}_{A,-1}\big), \\
i S_A^{[2]}& = & \frac{1}{U^2} \big(\left[{\cal K}_{A,+1},{\cal K}_{A,0}\right]
             +\left[{\cal K}_{A,-1},{\cal K}_{A,0}\right] \big),
\eea
which generalizes the expression in Eq.~(\ref{exprS}) to include the
vector potential. 

As explained in Appendix B, in the Monte Carlo calculation we treat the
canonical transformation as modifying the operator whose expectation
value is then taken in the fully projected BCS state; 
see Eq.~(\ref{transformed_op}). 

\vspace{0.1in}

\noindent{\bf Hamiltonian:}
Using the $i S_A^{[1]}$ derived above, it is easy to show that
the transformed Hamiltonian in the ${\cal D}=0$ sector is
given to ${\cal O}(t^2/U)$ by
\be
\tilde{\cal H}_A = {\cal K}_{A,0} + \frac{1}{U} \left[{\cal K}_{A,+1},
{\cal K}_{A,-1}\right].
\ee
For $A_{\br\brp}=0$ this reduces to the result of
Eq.~(\ref{tildeH}), while more generally we get 
Eq.~(\ref{transformed_H}) which was used in the derivation of
the optical spectral weight.

\vspace{0.1in}

\noindent{\bf Momentum Distribution:}
The momentum distribution $\la n_{\bk\sig} \ra$ is the
Fourier transform
of $\la {\cal G}_\sig(\br,\brp) \ra \equiv \la c^\dag_{\br\sig}
c_{\brp\sig} \ra$.
In parallel with our earlier notation for ${\cal K}$, we may write
the operator ${\cal G}_\sig(\br,\brp) =
{\cal G}_0 + {\cal G}_{+1} + {\cal G}_{-1}$, where ${\cal G}_n$ connects
the sector ${\cal D}$ to ${\cal D}+n$. 
The transformed operator $\tilde{\cal G} = \exp(iS){\cal G}\exp(-iS)$ 
to first order in $t/U$, with $A_{\br\brp}=0$, is given by 
\be
\tilde{\cal G}_\sig(\br,\brp) = {\cal G}_0(\br,\brp,\sig) -
\frac{1}{U} \big( {\cal K}_{-1} {\cal G}_{+1} + {\cal G}_{-1} {\cal K}_{+1}
\big).
\label{nk0}
\ee

Writing this explicitly in terms of electronic operators, with 
$h_{\br\sig} = (1-n_{\br\sig})$ and $\sigb = -\sig$, we get 
\bea
\tilde{\cal G}_\sig(\br,\brp) & = & 
h_{\br\sigb} c^\dag_{\br\sig} c_{\brp\sig} h_{\brp\sigb} \nn \\
&+& \frac{1}{U} \sum_{\bR,\sigp}
\left( t_{\br\bR} h_{\bR\sigpb} c^\dag_{\bR\sigp} c_{\br\sigp} n_{\br\sigpb}
c^\dag_{\br\sig} c_{\brp\sig} h_{\brp\sigb} \right. \nn \\
&+&  \left. t_{\brp\bR} h_{\br\sigb} c^\dag_{\br\sig} c_{\brp\sig} 
n_{\brp\sigb}
c^\dag_{\brp\sigp} c_{\bR\sigp} h_{\bR\sigpb} \right).
\label{nk1}
\eea
Note that the difference between the large U Hubbard and tJ model
momentum distributions shown in Fig.~\ref{figdiagnk} comes entirely from
the ${\cal O}(t/U)$ terms in Eq.~(\ref{nk1}), which would be
omitted in calculating $n(\bk)$ for the tJ model.

\vspace{0.1in}

\noindent{\bf First Moment of the Spectral Function:}
The first moment $M_1(\bk)$ is given by
\bea
M_1(\bk) &=& \la 0 \vert c^\dag_{\bk\sig} \big[ c_{\bk\sig},{\cal H}
\big] \vert 0 \ra - \mu \la n_{\bk} \ra \nn \\
&=& \big(\epsilon(\bk) - \mu\big) \la n_{\bk\sig} \ra
+ \Omega(\bk)
\eea
where $\Omega(\bk)$ is the Fourier transform of
\be
\Omega(\br,\brp) = U \la 0 \vert c^\dag_{\br\sig} c_{\brp\sig}
n_{\brp\sigb} \vert 0 \ra
\ee
Since $\Omega$ is of ${\cal O}(U)$, we have to canonically transform
it upto to second order in $t/U$ to get the moment $M_1(\bk)$ correct
upto ${\cal O}(J)$. Thus,
writing $\Omega(\br,\brp) = \Omega_0 + \Omega_{-1} + \Omega_{+1}$, we
find $\Omega_{+1} = 0$, and in the sector with ${\cal D}=0$,
\bea
\tilde{\Omega}(\br,\brp)&=&-\frac{1}{U} \Omega_{-1} {\cal K}_{+1}
+ \frac{1}{U^2} \Omega_{-1}\left[{\cal K}_0,{\cal K}_1\right] \nn\\
&+& \frac{1}{U^2} {\cal K}_{-1}\Omega_0 {\cal K}_1
\label{m1k}
\eea
The explicit expression for $\tilde{\Omega}(\br,\brp)$ in terms of
electron operators is rather lengthy and omitted for simplicity.

\bigskip

\section{Technical details of the Monte Carlo calculations}
\bigskip

The use of Monte Carlo methods in variational calculations
has a long history \cite{ceperley} and there have been many applications
to Hubbard and tJ models which are referenced in the text.
In this Appendix we discuss various technical points of the Monte Carlo
calculation, including (a) the choice of lattice and boundary conditions,
(b) the Monte Carlo moves in the sampling and their implementation, 
(c) details about number of configurations sampled for equilibration and 
for averaging data, and (d) various checks on our code.

To implement the Monte Carlo for evaluating expectation values of
operators on our wavefunction, we find it convenient to work with the
fully projected wavefunction ${\cal P} \vert\Psi_{\rm BCS}\ra$ and 
canonically transformed operators $\tilde{Q} = \exp(iS){\cal Q}\exp(-iS)$. 
At discussed in Sec.~III, this is equivalent to evaluating expectation 
values of ${\cal Q}$ in $\vert\Psi_0\ra$; see Eq.~(\ref{transformed_op}). 

\bigskip

\ni{\bf Lattice and Boundary Conditions:}

The BCS part of the variational wavefunction is written in 
coordinate space as a Slater determinant of pairs as shown in Eq.~(\ref{det}).
Each element of this determinant $\varphi(\br_{i\up}-\br_{j\dn})$
is the Fourier transform of $\varphi(\bk) = v_\bk/u_\bk$ defined below 
Eq.~(\ref{wavefn}). For a d-wave state $\tDelta_\bk = 0$ on 
the Brillouin zone (BZ) diagonals, which leads to a singularity
in $\varphi(\bk)$ at all $\bk$-points 
$\vert k_x\vert = \vert k_y\vert$ with 
$\epsilon(\bk) - \mu_{\rm var}\leq 0$.
For a numerical calculation it is thus best to avoid these $\bk$-points
by appropriate choice of the lattice and boundary conditions.
Three possible alternatives are: (1) a square lattice
with periodic/antiperiodic boundary conditions; or (2) a rectangular
lattice whose dimensions are mutually co-prime and periodic
boundary conditions (PBC); or (3) a ``tilted'' lattice, described further 
below, with PBCs. All three schemes lead to a set of $\bk$-points which
avoid the zone diagonal on any finite system.

We have chosen to work on a tilted lattice even though it is perhaps the
least intuitively obvious of the three alternatives because it
preserves the four-fold rotational symmetry of the lattice and also
does not introduce any twists in the boundary conditions
(which might be important in a state with long range SC order).
We have later checked that our results for doped systems
($x>0$) are not dependent on this
choice by comparing them with option (1).

\begin{figure}
\begin{center}
\epsfig{file=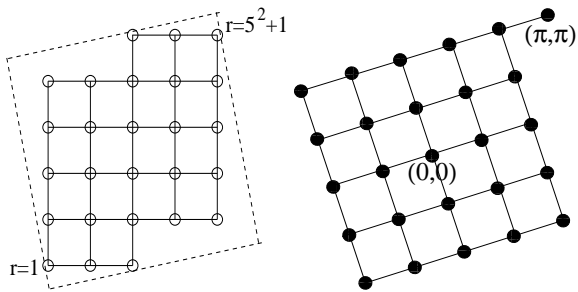,width=3.0in,angle=0}
\vskip5mm
\caption{{\bf Left}: Real space picture of the $L^2+1$ lattice for $L=5$,
with periodic
boundary conditions applied along the opposite edges of the tilted
square indicated by dashed lines. {\bf Right}: The $\bk$-space Brillouin zone
of the ``tilted lattice'' for $L=5$. In the calculations reported in this paper
we used systems with $L = 15, 17, 19$.}
\label{figlattice}
\end{center}
\end{figure}

The tilted lattice with PBCs was also used in the early work of
Gros and coworkers \cite{gros87,gros88}. These lattices have
$L^2+1$ sites with odd $L$; an example with $L=5$ is 
shown in Fig.~\ref{figlattice}(a). The corresponding BZ is a tilted
square of allowed $\bk$-points shown in Fig.~\ref{figlattice}(b).
More generally, the allowed $\bk$-points are the solutions of
$\exp(i k_x L + i k_y) = 1$ and $\exp(i k_y L - ik_x) = 1$.
This leads to the $(L^2 + 1)$ solutions:
$k_x = 2\pi (mL - n)/[L^2 + 1]$ and
$k_y = 2\pi (m + nL)/[L^2 + 1]$ with
$m = -(L-1)/2, \ldots, +(L-1)2$, 
$n = -(L-1)/2, \ldots, +(L-1)2$ and
the single additional point $\bk = (\pi,\pi)$
corresponding to
$(m,n) = \left((L+1)/2,(L-1)/2\right)$. 

Note that the $\bk=0$ point is not avoided in this scheme,
and we choose $\varphi(\bk=0)$ to be a very large but
finite number, and check that we recover standard BCS results
independent of this choice. Further checks of our procedure are
described below.

\bigskip

\ni{\bf Monte Carlo method:}

To sample configurations for evaluating expectation values, we use the
standard variational Monte Carlo method using the Metropolis
algorithm to generate a sequence of many-body configurations
distributed according to 
$\left\vert {\cal P}
\langle \{\br_i\},\{\brp_j\} \vert \Psi_{\rm BCS} \rangle
\right\vert^2$.
The Monte Carlo moves used are:
(i) choosing an electron and moving it to an empty site
and (ii) exchanging two antiparallel spins.
Starting in the ${\cal D}=0$ sector these moves conserve ${\cal D}$;
thus the no double-occupancy constraint ${\cal P}$
is trivial to implement exactly.
Also all allowed states in the ${\cal D}=0$ sector (with $S_z^{\rm
tot}=0$) can be accessed.
For an $N$-electron system, the moves involve updating
the determinant of the $\frac{N}{2}\times\frac{N}{2}$ matrix of
Eq.~(\ref{det}). We do this using the inverse update method of 
Ceperley, Chester and Kalos \cite{ceperley},
the time for which scales $\sim N^2$, in contrast to $\sim N^3$ for
directly evaluating the determinant of an updated configuration.

\bigskip

\ni{\bf Numerical Details:}

Much of the data were obtained on $L=15$ ($L^2 + 1 = 226$-site) lattices. Some
runs were on an $L=19$ ($362$-site) lattice to reduce finite size errors
on the order parameter at overdoping and for better $\bk$-resolution
for $n(\bk)$. We equilibrated the system for about $5000$ sweeps, where
every electron is updated once on average per sweep. Typically we
averaged data over $1000$ configurations chosen from about 
$5000$ sweeps. For some parameter values we performed
long runs of $10^5$ sweeps. Specifically, such long runs were used to
calculate quantities such as the order parameter at 
at certain doping values to reduce statistical error bars.
In most figures the error bars are not explicitly shown, because
the errors from the stochastic Monte Carlo evaluation are smaller 
than the symbol size. 

\bigskip

\ni{\bf Checks on the code:}

To check our code we have made detailed comparisons 
against published results of the ground state energy
\cite{gros88} for appropriate parameter values.
At various points in the text we also mentioned other checks
we have made on the limiting values of several observables.
We have checked that in the 
low electron density (nearly empty lattice) limit,
the quasiparticle weight $Z \to 1$, our estimated
$v_{_F}$ approaches the bare Fermi velocity, and the total
optical spectral weight vanishes as $x \to 1$.

Here we describe three additional checks we have made in the $x=0$
insulating limit.
First, it is well-known that at $x=0$ the canonically
transformed Hamiltonian $\tilde{\cal H}$
can be rewritten as the Heisenberg spin model
\be
{\cal H}_{\rm AF} = \sum_{\br\brp} J_{\br\brp} \left( \bS_\br \cdot
\bS_{\brp} - \frac{1}{4}\right)
\label{Hafm}
\ee
where $J_{\br\brp} = 4 t^2_{\br\brp}/U$. 
We can thus compute the ground state energy in two
different ways: either by directly using $\tilde{\cal H}$
from Eq.~(\ref{tildeH}), 
or by calculating the ground state spin correlations 
$\la\bS_\br \cdot \bS_{\brp}\ra =3\la S^z_\br S^z_{\brp}\ra$
(from spin rotational invariance in the singlet ground state) and
then using Eq.~(\ref{Hafm}) to get the energy.
We have verified that these two estimates agree 
$\la{\tilde{\cal H}}\ra = \la{\cal H}_{\rm AF}\ra$,
which serves as a nontrivial check on our code.
Note that, unlike in the rest of the paper, in the remainder of
this Appendix we use the symbol $\la \ldots \ra$ to mean the
expectation value in the state ${\cal P}\vert \Psi_{\rm BCS} \ra$,
{\em without} the factor of $\exp(-iS)$.

Second, the canonically transformed Fourier transform of the momentum
distribution, $\tilde{\cal G}_\sig({\br,\brp})$ in Eq.~(\ref{nk1}),
may be related to spin correlations at $x=0$ 
(see also Refs.~\cite{takahashi77,eskes94,oganesyan00}) as
\be
\tilde{\cal G}_\sig({\br,\brp}) = 2 \frac{t_{\br\brp}}{U}\big(\frac{1}{4} -
\bS_\br\cdot\bS_{\brp}\big).
\label{nkspin}
\ee
We have explicitly checked our code by 
calculating $\la{\tilde{\cal G}}\ra$ 
from Eq.~(\ref{nk1}) and independently evaluating 
the spin correlation $\la \bS_\br\cdot
\bS_{\brp} \ra$, and verifying the relation in Eq.~(\ref{nkspin}).

Finally, for the first moment calculation described at the end of
Appendix A we find the following simple result at $x=0$. 
For the cases $\brp=\br$ and $\brp\neq\br$, we find respectively
\be
{\tilde\Omega}(\br,\br)=\frac{2}{U} \sum_{\bR} t_{\br\bR}^2 S_{\br\bR}
\ee
\be
{\tilde\Omega}(\br,\brp)\!\!=\!\!t_{\br\brp} S_{\br\brp} +
\sum_{\bR} \frac{t_{\br\bR} t_{\brp\bR}}{2 U} \left(S_{\brp\bR}+
S_{\br\bR}- S_{\br\brp} \right)
\label{m1spin}
\ee
with $S_{\br\brp} \equiv \left(1/4 - \la \bS_{\br}\cdot\bS_{\brp} \ra
\right)$. 
We have verified that the moments computed directly using Eq.~(\ref{m1k})
agree with those obtained using expressions above in terms of spin
correlation functions, which serves as yet another nontrivial check.

\bigskip
\section{Slave boson mean field theory}
\bigskip

In this Appendix we first briefly summarize the results of
slave-boson mean-field theory (SBMFT) for the tJ model
\cite{bza87,kotliar88} and then compare them
with the variational results presented in the text.
Many authors have used SBMFT with small
variations and it is important to unambiguously define our
notation to make detailed comparisons.

\medskip
The tJ model is defined by the Hamiltonian in Eq.~(\ref{HtJ})
acting on the Hilbert space with
$\sum_{\sig} n_{\br\sig} \leq 1$ at each site $\br$.
Following standard slave-boson methodology
\cite{kotliar88}, we can write
$c_{\br\alpha}^{\dag} = b_\br f_{\br\alpha}^{\dag}$
where $f_{\br\alpha}^\dag$ creates a neutral spin-$1/2$ fermion
(spinon) and $b^\dag_{\br}$ a spinless charge-$e$ boson (holon).
The constraint at each site is now:
$\sum_\alpha f_{\br\alpha}^\dag f_{\br\alpha} + b_{\br}^\dag
b_{\br} = 1$. 
The Hamiltonian can now be written as
\bea
{\cal H}_{\rm tJ}&=&- \sum_{\br,\brp,\sig} t_{\br\brp}
b_{\br} f_{\br\sig}^{\dag} f_{\brp\sig} b_{\brp}^\dag \\
&+& J \
\sum_{\la\br\brp\ra} \big[ \bS^f(\br)\cdot\bS^f(\brp)
- \frac{1}{4} (1-b_\br^\dag b_\br)(1-b_{\brp}^\dag b_{\brp}) \big] \nn
\eea
Here $t_{\br\brp}=t$ for nearest neighbors, and $(- t^\prime)$ for
next-nearest neighbors, which fixes the bare dispersion $\epsilon(\bk)
=-2 t (\cos k_x + \cos k_y) + 4 t^\prime \cos k_x \cos k_y$.
The next-nearest neighbor $J^\prime/J = 1/16$ is ignored.

Following Ref.~\cite{kotliar88} we make three approximations.
First, we make a Hartree-Fock-Bogoliubov mean-field approximation for the
$\bS^f(\br)\cdot\bS^f(\brp)$ term. Second, we assume that the bosons 
are fully condensed at $T=0$ so that $\la b \ra = \sqrt{x}$. 
Third, we make the (most drastic) approximation that the constraint
is obeyed {\em on average} and {\em not necessarily at each site}.
This leads to the mean-field Hamiltonian
\be
{\cal H}_{\rm MF} = \sum_{\bk\sig} \big[\tilde{\epsilon}(\bk)- \tilde{\mu}
\big]
n_{\bk\sig} + \sum_{\bk} \Delta_\bk \big( f_{\bk\up}^\dag f_{-\bk\dn}^\dag
+ {\rm h.c.} \big)
\label{MFH}
\ee
where $\tilde{\epsilon}(\bk)= -2 (x t + 3 J \chi/4) (\cos k_x + \cos
k_y) + 4 x t' \cos k_x \cos k_y$, and $\Delta_\bk = \Delta^{\rm sb} (\cos
k_x - \cos k_y)/2$. The pairing field, $\Delta^{\rm sb} = 3 J \vert \la
f_{\br\up}^\dag f_{\brp\dn}\ra\vert$, the
Fock field $\chi = \la f_{\br\sig}^\dag f_{\brp\sig}\ra$, and the
``chemical potential'' $\tilde{\mu}$, are determined through the
following set of self-consistent equations:
\bea
\frac{1}{J}&=&\frac{3}{8} \int \frac{d^2\bk}{(2\pi)^2}
\frac{(\cos k_x - \cos k_y)^2}{E_\bk} \\
\chi&=&-\frac{1}{4} \int \frac{d^2\bk}{(2\pi)^2}
\big(\frac{\xi_{\bk}}{E_{\bk}}\big) (\cos k_x + \cos k_y) \nn \\
x&=& \int
\frac{d^2\bk}{(2\pi)^2}\big(\frac{\xi_{\bk}}{E_{\bk}}\big)\nn 
\eea
where $\xi_\bk = \tilde{\epsilon}(\bk)-\tilde{\mu}$ and
$E_\bk = \sqrt{\xi_\bk^2 + \Delta^2_\bk}$.

These equations can be numerically solved 
and the results summarized as follows: (i) $\chi(x)$ and
$\tilde{\mu}(x)$ are smooth non-singular functions. In the insulator,
$\chi(0)\neq 0$, leading to a finite spinon dispersion determined by
$J$. (ii) $\Delta^{\rm sb}(x=0)$ is finite at $x=0$ and its scale
is determined purely by $J$. (iii) $\Delta^{\rm sb}(x)$ decreases with
increasing doping, vanishing at a critical $x=x_c \approx 0.35$-$0.4$, 
which is a weak function of $J$.



We next calculate various physical quantities within SBMFT
and compare with our variational results.
The SC order parameter is given by
$\Phi^{\rm sb} \equiv \vert \la c_{\br\up}
c_{\br+\delta\dn} \ra\vert = x \Delta^{\rm sb}(x)/3 J$
where the explicit factor of $x$
comes from $\vert \la b \ra \vert^2 = x$. 
The SBMFT thus correctly captures the nonmonotonic behavior of $\Phi$, 
vanishing in the limits $x \to 0$ and $x\to x^{-}_c$, and maximum at 
$x \sim 0.15$-$0.2$. In the language of SBMFT, although the
spinons are paired, the order parameter $\Phi^{\rm sb}(x=0)=0$ since
there are no holons to condense. On the other hand,
the SC-Fermi liquid transition on the overdoped side corresponds to 
the vanishing of the spinon pairing amplitude $\Delta^{\rm sb}$.

In Sec.~VIII we compared SBMFT results for the nodal
quasiparticle weight and dispersion with the corresponding 
variational results. 
Although the SBMFT language of spinons and holons appears to be very
appealing, it would be justified only if the spinons and holons were
essentially non-interacting particles, at least at sufficiently low
energies. Our conclusion in Sec.~VII was that this is not the case
and the approximation of decoupling the holon and spinon Greens functions 
is not valid in computing, e.g., the nodal quasiparticle residue Z;
see Sec.~VIIIA.

Here we give sketch the derivation of these SBMFT results.
Within SBMFT the electron Green function factorizes
to give $G(\bk,\omega) = x G_f(\bk,\omega)$, where $x$ comes from 
the condensed holons and $G_f$ is the spinon Green function obtained
from ${\cal H}_{\rm MF}$ in Eq.~(\ref{MFH}) above. 
Note that SBMFT does not capture the incoherent part of the spectral function 
and also does not satisfy sum rules.
This factorization
leads to the
following results for the nodal quasiparticles within SBMFT: (i)
Along the zone diagonal the spinon $n_f(\bk) = \theta(-\xi_\bk)$
and thus the nodal quasiparticle residue $Z^{\rm sb} = x$. 
Thus $Z^{\rm sb} < Z$, where $Z$ is the variational estimate
(see Fig.~\ref{figdiagnk}(c)), and this inequality implies the
inadequacy of the spinon-holon decoupling.
(ii) The quasiparticle dispersion is obtained from
the poles of $G(\bk,\omega)$ and this is entirely
governed by the spinon dispersion. 
At low doping, the SBMFT $v^{\rm sb}_{_F}(x)=3 J \chi+4 x t$ 
and is smaller than the
variational estimate $v_{_F}(x)$ (see Fig.~\ref{fignodal}(b)) 
and also exhibits much more doping dependence. 
Despite large quantitative differences, there is
one important qualitative similarity: both
$v^{\rm sb}_{_F}(x)$ and $v_{_F}$ go to a non-zero limit
as $x \to 0$.


\begin{thebibliography}{99}
\bibitem{BCS_footnote}
The connection with the usual ``fixed phase'' BCS wave function can be
seen as follows. Ignoring normalization factors, we can see that
$\vert \Psi_{\rm BCS} \rangle$ can be obtained as the $N$-particle projection
of $\exp\left(\sum_\bk \varphi(\bk)
c_{\bk\uparrow}^{\dag}c_{-\bk\downarrow}^{\dag}\right) \vert 0 \rangle$.
This in turn can be rewritten as
$\prod_\bk \exp\left(\varphi(\bk)
c_{\bk\uparrow}^{\dag}c_{-\bk\downarrow}^{\dag}\right) \vert 0 \rangle 
= \prod_\bk \left(1 + \varphi(\bk)
c_{\bk\uparrow}^{\dag}c_{-\bk\downarrow}^{\dag}\right) \vert 0 \rangle$,
which is clearly the same as the original BCS state.
\bibitem{footnote_z}
The nodal $k_F$ is determined from the location of
the slope discontinuity in the first moment $M_1(\bk)$ along the
zone diagonal as discussed in Sec.~VIII B. This estimate,
which turns out to be quite close to the non-interacting $k_F$, 
is a much more accurate way of determining $k_F$ than
obtaining it from $n(\bk)$ given the finite $\bk$-resolution
Next polynomial fits are made to the numerical data points
for $n(\bk)$ for $k > k_F$ and for $k < k_F$. The intersections
of the fit curves with vertical line at $k = k_F$ are then used to
determine $Z$.
\bibitem{footnote.oldpaper}
In our earlier Letter\cite{paramekanti01}, we had compared $t\tilde{\Delta}_{\rm var}
= \Delta_var(x)$ directly with the hump scale (and for our choice of
parameters, $t = 3 J$). However, we believe that the energy scale for pairing
is actually determined by $J$ (and not $t$) as discussed in the text.
\bibitem{footnote.units}
In our earlier Letter\cite{paramekanti01}, we had inadvertently
used the in-plane lattice spacing $a$ rather than the c-axis lattice
spacing $d$ to convert the computed $D_{\rm low}$ to $\omega^\ast_p$. 
This did not affect any of the conclusions of the earlier paper,
where the primary interest was in the doping dependence of $D_{\rm low}$.
\end{thebibliography}
\end{document}